\documentclass[pre,aps,twocolumn,superscriptaddress]{revtex4}

\usepackage{dcolumn}
\usepackage{bm}
\usepackage{amssymb}
\usepackage{color}
\usepackage[pdftex]{graphicx}

\usepackage{amsmath}
\usepackage[linesnumbered,ruled]{algorithm2e}

\usepackage{url}
\usepackage[section]{placeins}
\usepackage[colorlinks=true,linkcolor=blue,citecolor=blue]{hyperref}

\begin{document}


\title{Cell dynamics simulations of coupled charge and magnetic phase transformation \\ in correlated oxides}

\author{Lingnan Shen}
\affiliation{Department of Physics, University of Virginia, Charlottesville, VA 22904, USA}


\author{Gia-Wei Chern}
\affiliation{Department of Physics, University of Virginia, Charlottesville, VA 22904, USA}

\date{\today}

\begin{abstract}
We present a comprehensive numerical study on the kinetics of phase transition that is characterized by two non-conserved scalar order parameters coupled by a special linear-quadratic interaction. This particular Ginzburg-Landau theory has been proposed to describe the coupled charge- and magnetic transition in nickelates and the collinear stripe phase in cuprates. The inhomogeneous state of such systems at low temperatures consists of magnetic domains separated by quasi-metallic domain-walls where the charge-order is reduced. By performing large-scale cell dynamics simulations, we find a two-stage phase-ordering process in which a short period of independent evolution of the two order parameters is followed by a correlated coarsening process. The long-time growth and coarsening of magnetic domains is shown to follow the Allen-Cahn power law. We further show that the nucleation-and-growth dynamics during phase transformation to the ordered states is well described by the Kolmogorov-Johnson-Mehl-Avrami theory in two dimensions. On the other hand, the presence of quasi-metallic magnetic domain walls in the ordered states gives rise to a very different kinetics for transformation to the high temperature paramagnetic phase. In this new scenario, the phase transformation is initiated by the decay of magnetic domain walls into two insulator-metal boundaries, which subsequently move away from each other. Implications of our findings to recent nano-imaging experiments on nickelates are also discussed. 
\end{abstract}

\maketitle

\section{Introduction}
\label{sec:intro}

Metal insulator transition in correlated electron materials is a complex process which often involves multiple order parameters~\cite{imada98,dobrosavljevic12}. In particular, almost all such electronic phase transitions are accompanied by structural distortions. Other degrees of freedom such as spins and orbitals also often play an important role in the transformation from metal to insulating phases, and vice versa. The intricate interplay and competition between different ordering tendencies make the modeling of metal-insulator transition a very difficult task. Moreover, several recent nano-imaging experiments have revealed a highly inhomogeneous process for such phase transformations in correlated materials~\cite{atkin12,liu17,qazilbash07,lupi10,mcleod16,mattoni16,preziosi18,li19,post18}. 
A complete theoretical description of the metal-insulator transition dynamics thus requires theoretical efforts across multiple spatial and time scales.

Macroscopically, the phase-field method has been widely used to investigate phase separation, pattern formation, and nucleation-and-growth phenomena in complex systems~\cite{binder73,valls90,puri09,onuki02,chen02}. The phase field approach is intimately related to the concept of order parameter for characterizing the different symmetry-breaking phases. Dynamical models of the order-parameter fields have also been systematically studied and classified~\cite{bray02,hohenberg77}. Notable among them are the time-dependent Ginzburg-Landau dynamics for non-conserved order parameter field, and the Cahn-Hilliard dynamics for conserved order parameters~\cite{puri09,onuki02}.

It is worth noting that Mott transition in Hubbard-type models describes a transition between different dynamical regimes: the metal and insulator phases correspond microscopically to regimes of itinerant and localized electrons, respectively. The ``ideal" Mott transition is thus not associated with broken symmetries as in conventional phase transitions, although, as mentioned above, metal-insulator transitions in real materials are often accompanied by structural or magnetic orderings.  Nonetheless, even for the ideal Mott transition, a scalar order parameter characterizing a broken Ising-type symmetry is often used to describe the correlation-induced electron localization based on the similarity with the liquid-gas transitions~\cite{limelette03,papanikolaou08,bar18,kundu20}. More specifically, the scalar order parameter can be defined as the local density of doubly occupied sites or doublons~\cite{kotliar86,wang10,sandri13,chern19}. As in the liquid-gas case, most Mott metal-insulator transitions are discontinuous first-order transitions, underscoring the nonequilibrium character of the corresponding phase-transition kinetics. 

Interestingly, a recent nano-infrared imaging experiment uncovered intriguing concurrent first- and second-order electronic transitions in epitaxial NdNiO$_3$ films~\cite{post18}. By directly probing the local electronic conductivity, the nano-IR images showed competing metallic and insulating domains co-existing in the sample bulk, a characteristic of first-order phase transition. On the other hand, conducting domain walls in a temperature range proximal to the first-order transition are found to exhibit an optical conductivity that evolves gradually from insulating to metallic as the sample is warmed. These anomalous nanoscale domain walls are ascribed to boundaries between different antiferromagnetically ordered regions within the charge-ordered bulk. 

The complex behaviors observed in nickelates can be understood from a Landau theory of coupled charge and magnetic orders~\cite{lee11,peil19,post18}, highlighting the importance of coupled order parameters in driving the metal-insulator transitions in correlated oxides. Importantly, while both the charge and magnetic orders are described by a non-conserved scalar order parameter, a unique linear-quadratic coupling term allowed by symmetry renders the transition a discontinuous one. It is worth pointing out that similar Landau theory, in which two coupled order parameters are introduced to describe the intertwined charge and spin-density waves ordering, has previously been proposed to describe the stripe phases in cuprates and nickelates~\cite{zachar98}. 

Although this Landau theory has been shown to successfully explain the possible phases and the domain-wall structures in the ordered states~\cite{post18}, the dynamical behaviors of the coupled charge and spin systems, especially nonequilibrium phase transformation dynamics, has yet to be carefully investigated. In particular, it is unclear how the special linear-quadratic coupling affects the coarsening, nucleation and domain growth processes.  In this paper, we perform the first systematic and comprehensive investigations of the nonequilibrium phase-ordering dynamics in such systems based on large-scale cell dynamics simulations. 

The rest of the paper is organized as follows. In Sec.~\ref{sec:landau}, we outline the Landau theory, and the dynamical model used in this work. We also review the cell dynamics method for simulating the coupled time-dependent Ginzburg-Landau equations. We discuss the phase-ordering, or coarsening, dynamics of the coupled charge and magnetic orders in Sec.~\ref{sec:coarsening}. While our large-scale simulations confirm the scaling behavior expected for non-conserved order parameters, we find an interesting two-stage phase-ordering process. In Sec.~\ref{sec:nucleation}, we present extensive simulations of nucleation-and-growth phenomena for quenches to the coexisting regime. We show that the kinetics of the transformation into the ordered phase is well described by the Kolmogorov-Johnson-Mehl-Avrami theory. On the other hand, a rather unusual dynamics is found for transition to the high-temperature paramagnetic phase due to the pre-existing quasi-metallic domain-walls. Finally, we summarize our results and discuss their experimental implications in  Sec.~\ref{sec:summary}. 

\section{Phase field model and cell dynamics simulations}
\label{sec:landau}

\subsection{Landau theory of coupled charge and spin order}

The ordered phases of correlated nickelates $R$NiO$_3$, where $R$ is a rare-earth element, are characterized by a charge order with a doubled unit cell, and a unique antiferromagnetic order with quadrupled unit cell. Specifically, the wave vectors of these two orderings are $\mathbf k_{\rm CO} = [\frac{1}{2}, \frac{1}{2}, \frac{1}{2}]$ and $\mathbf k_{\rm AF} = [\frac{1}{4}, \frac{1}{4}, \frac{1}{4}]$, respectively, in the pseudo-cubic lattice setting~\cite{post18}. For $R = $ Nd or Pr, the transition into the low-$T$ phase occurs through a single first-order transition where the charge and antiferromagnetic orders emerge simultaneously~\cite{torrance92,georgescu19}. On the other hand, for $R$ smaller than Nd, the two orderings take place at different temperatures, with charge-ordering appearing at a higher temperature~\cite{torrance92}. Moreover, the two separated transitions are both second-order. These two distinct phase-transition scenarios highlight the nontrivial interplay between the charge and spin order parameters.

\begin{figure}[t]
\includegraphics[width=0.99\columnwidth]{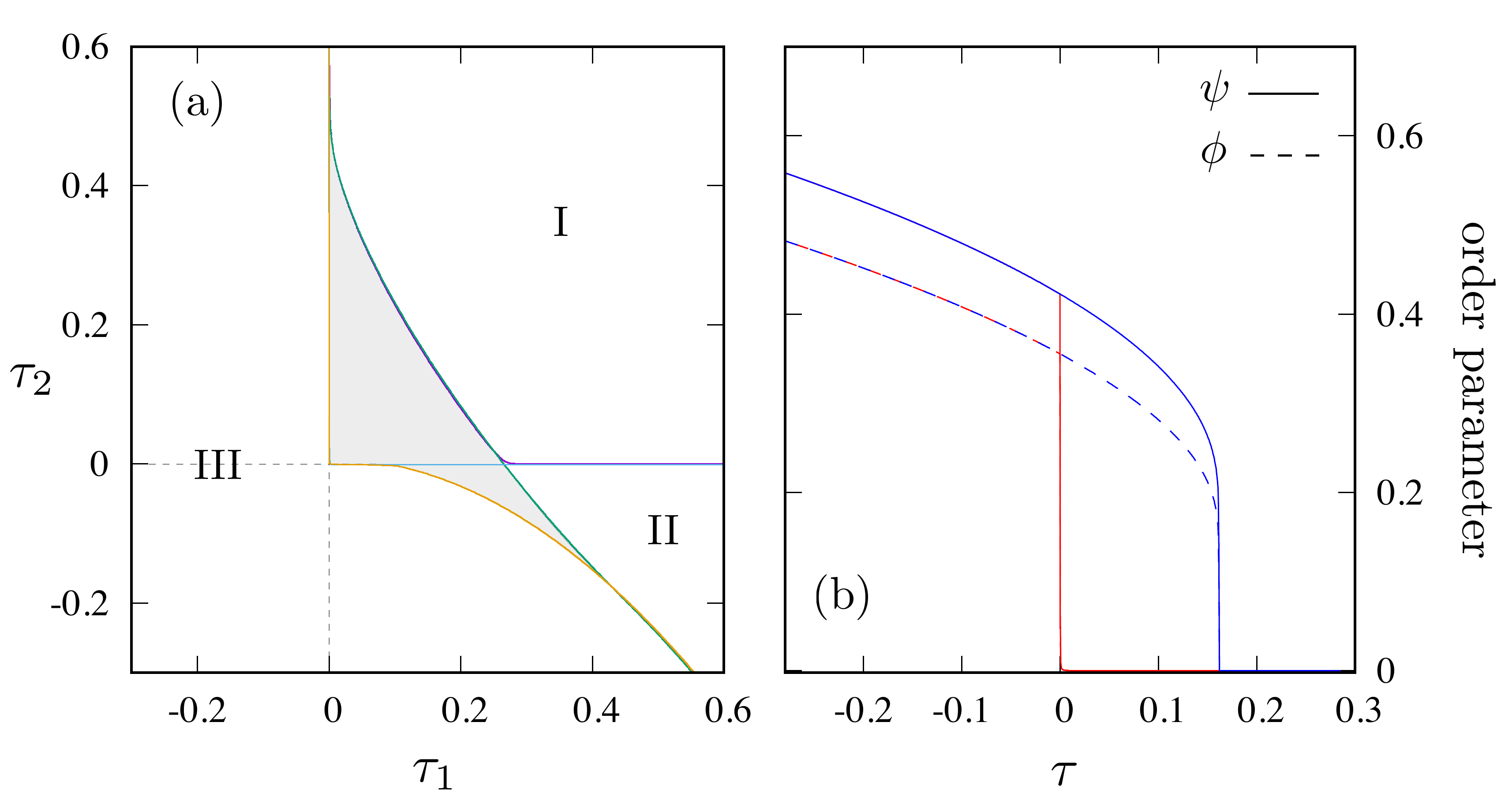}
\caption{(Color online)  
\label{fig:phase} (a) Phase diagram of the Landau theory given by Eq.~(\ref{eq:landau}) with parameters $g_1 = g_2 = 1$ and $\lambda = 2$. The various phases are: paramagnetic phase (I), charge-ordered phase without magnetic order (II), and simultaneous charge- and spin-ordered state (III). The shaded area indicates co-existing regime of the first-order phase transition. (b) the charge ($\phi$) and spin ($\psi$) order parameter as a function of reduced temperature $\tau$ along the diagonal line $\tau_1 = \tau_2$, showing a clear hysteresis loop characteristic of a first-order transition. 
}
\end{figure}

Following Ref.~\cite{post18}, we introduce two scalar fields $\phi$ and $\psi$ to describe the charge and magnetic ordering, respectively. In general, due to the Ising-type $Z_2$ symmetry of such scalar order parameters, the leading-order coupling between them in the Landau free-energy expansion is a biquadratic term $\psi^2\, \phi^2$. Such interaction term has been shown to induce novel magneto-electric coupling present at the domain walls~\cite{daraktchiev81}. However, this biquadratic coupling cannot induce a first-order transition for the two scalar order parameters. On the other hand, the fact that the propagation vector of the charge order is commensurate with that of spin order through $\mathbf k_{\phi} = 2 \mathbf k_{\psi}$ indicates that a unique linear-quadartic coupling is allowed by symmetry~\cite{lee11,post18}. The Ginzburg-Landau free-energy for such a system is then given by
\begin{eqnarray}
	\label{eq:landau}
	\mathcal{F}[\phi, \psi] &=& \frac{1 }{2}\left(\nabla \phi\right)^2 + \frac{ \tau_{1}}{2} \,\phi^2 + \frac{ g_{1}}{4} \phi^4 \nonumber \\
	&+ &  \frac{1}{2} \left(\nabla \psi\right)^2 + \frac{ \tau_2 }{2} \psi^2 +\frac{ g_{2}}{4} \psi^4 - \lambda \psi^2 \phi,
\end{eqnarray}
where $\tau_j = r_j (T - T^c_j)/T^c_j$ with $j = 1, 2$ are the reduced temperatures, $r_j$ are positive constants. The coefficients~$g_j$ of the quartic terms, as well as the linear-quadratic coupling $\lambda$ are assumed to depend weakly on temperature. For simplicity, we normalize the two order parameters such that the coefficients of the gradient term are dimensionless. The phase diagram of this model, obtained by minimization of $\mathcal{F}$ with respect to $\psi$ and $\phi$, is shown in Fig.~\ref{fig:phase}(a). At high temperatures, both order parameters vanish $\phi = \psi = 0$, corresponding to disordered charges and spins in the paramagnetic phase, labeled as phase~I in the phase diagram. In the absence of the $\lambda$ term, the two order parameters are decoupled, and long-range ordering of charge and spins proceeds through a second-order phase transition independently when the respective~$\tau_j$ becomes negative at low temperatures.

Importantly, a nonzero spin-charge coupling $\lambda \neq 0$ changes the nature of phase transition to first-order in the vicinity of the multi-critical point $\tau_1 = \tau_2 = 0$, around which an extended region of coexistence (shaded area) appears; see Fig.~\ref{fig:phase}(a). The first-order transition also manifests itself in the pronounced hysteresis loop shown in Fig.~\ref{fig:phase}(b) as the system is annealed and warmed along the diagonal line $\tau_1 = \tau_2$. For negative $\tau_2 < 0$ and positive $\tau_1 \gg |\tau_2| > 0$, the symmetry-breaking phase, labeled by phase II in Fig.~\ref{fig:phase}, is characterized by $\phi > 0$ and $\psi = 0$, i.e. a charge order without magnetic order. The nonzero charge-order parameter gives rise to a shifted quadratic coefficient $\hat{\tau}_2 = \tau_2 -2 \lambda \phi$ for the spin order, which physically increases the critical temperature for magnetic phase transition.

On the other hand, the presence of spin-order $\psi \neq 0$ when $\tau_1 < 0$ immediately induces a concomitant charge-order in phase~III through the linear-quadratic coupling. It is worth noting that the $Z_2$ symmetry of the charge-order is explicitly broken by the linear-quadratic coupling~$\lambda$. However, a global $Z_2$ symmetry remains in the spin sector of the ordered phase. The domain walls that separate the two distinct antiferromagnetic regions gives rise to quasi-metallic strings embedded in the charge-ordered insulating state. 
We also note that the scalar field~$\phi$, while describing a unit-cell doubled charge order, also serves as the primary order parameter for metal-insulator transition. Specifically, the metallic and the insulating phases in this work correspond to $\phi > 0$ and $\phi \approx 0$, respectively.

The Landau theory in Eq.~(\ref{eq:landau}) has been first proposed to describe the stripe phase in lanthanum nickelate and cuprate families of doped antiferromagnets~\cite{zachar98}. In this context, the two scalar order parameters $\phi$ and $\psi$ represent the fundamental Fourier components of the charge density wave (CDW) and collinear spin density wave (SDW) ordering, respectively. The linear-quadratic coupling $\lambda$ is allowed since the period of the CDW is generically half that of the SDW. The results presented in this work thus can also be applied to understanding the phase-transition dynamics of collinear stripes in these compounds. 

Finally, the same Landau energy functional also appears in model C systems in which a non-conserved scalar order parameter is coupled to a conserved concentration field~\cite{kockelkoren02,das17}. Notable examples  include intermetallic alloys, adsorbed layers on solid substrates, and supercooled liquids. In such systems, the decomposition process described by the conserved field is coupled to the ordering process which is modeled by the scalar order parameter. However, the dynamics of such model C systems is very different from that of the coupled charge and spin ordering studied in this paper. Since the charge and magnetic orders are both described by non-conserved order parameters, they follow the  the TDGL equation.  On the other hand, the conserved field in the model C system is governed by the Cahn-Hilliard-Cook (CHC) equation while the scalar order parameter obeys the TDGL equation. Nonetheless, as will be shown later, the coarsening process in both cases share some similar features.

\subsection{Cell dynamics method}

For non-conserved order parameter fields such as the charge and spin order in our case, the dynamics of the phase transformation is described by the model-A or the stochastic time-dependent Ginzburg-Landau (TDGL) equation~\cite{bray02,hohenberg77,puri09}
\begin{subequations}
\label{eq:TDGL}
\begin{eqnarray}
	\frac{\partial \phi}{\partial t} &=& -\Gamma_1 \frac{\partial \mathcal{F}}{\partial \phi} + \sigma_1,  \\
	\frac{\partial \psi}{\partial t} &=& - \Gamma_2 \frac{\partial \mathcal{F}}{\partial \psi} + \sigma_2, 
\end{eqnarray}
\end{subequations}
where $\Gamma_j$ are phenomenological relaxation coefficients and the two $\sigma$ are Gaussian white noises satisfying the following expectation values 
\begin{eqnarray}
	 \langle \sigma_j(\mathbf r, t) \rangle &=& 0,  \\
	 \langle \sigma_i(\mathbf r, t) \sigma_j(\mathbf r', t') \rangle &=& 2  \Gamma_j k_B T\,  \delta_{ij} \,\delta(\mathbf r - \mathbf r') \delta(t - t'). \nonumber
\end{eqnarray}
Numerical approaches for solving such phase-field models include standard finite-difference approximation as well as spectral methods. Computational studies of both TDGL and CHC equations have elucidated many aspects of phase separation for a wide range of system. An alternative numerical method, inspired by the cellular automaton theory, is the cell dynamics simulation~\cite{oono87,oono88,puri88}. Indeed, cell dynamics can be viewed as a new form of discrete dynamics on lattices. Compared with other numerical approaches, cell dynamics is considerably more efficient in describing the phase separation and domain growth in many systems, where the dynamics is dominated by diffusive processes.

\begin{figure*}
\includegraphics[width=1.99\columnwidth]{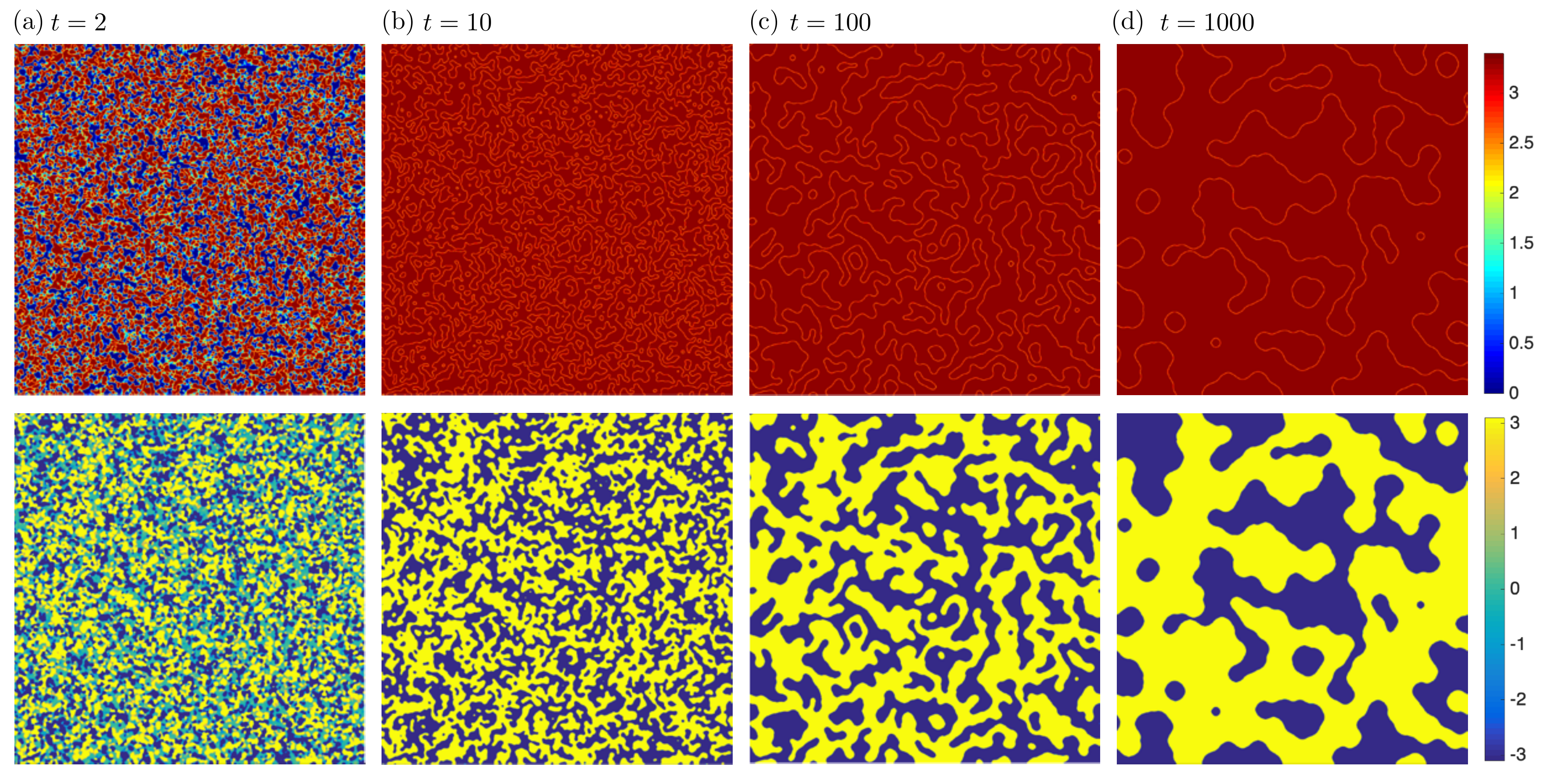}
\caption{(Color online)  
\label{fig:coarsening} Snapshots of coarsening in the ordered phase at a reduced temperature $\tau^* = -1.9$ obtained from cell-dynamics simulations on a $2000\times 2000$ square-lattice grid. The charge-spin coupling is $\lambda = 2$. The top and bottom panels show configurations of the charge $\phi$ and spin $\psi$ order parameters, respectively. The time-step used in the simulation is $\Delta t = 0.01$. 
}
\end{figure*}

In the cell dynamics method, the partial differential equation~(\ref{eq:TDGL}) is replaced by a finite difference equation in space and time in the following form
\begin{subequations}
\label{eq:cds_eq}
\begin{eqnarray}
	\phi_{\mathbf n}^{ t+ 1} &=& f_1[\phi_{\mathbf n}^t, \psi_{\mathbf n}^t] +
	 D_1 \left( \langle\langle \phi_{\mathbf n}^t \rangle\rangle - \phi_{\mathbf n}^t \right)   +  \eta_{1, \mathbf n}^t , \quad \\
	\psi_{\mathbf n}^{ t+ 1} &=& f_2[\phi_{\mathbf n}^t , \psi_{\mathbf n}^t ] + 
	D_2 \left( \langle\langle \psi_{\mathbf n}^t \rangle\rangle - \psi_{\mathbf n}^t \right)   +  \eta_{2, \mathbf n}^t. \quad
\end{eqnarray}
\end{subequations}
Here the first term represents the flow due to a one-to-one map $f : \mathbb{R} \to \mathbb{R}$, and the second term, with $\langle\langle * \rangle\rangle$ denoting an isotropic local average over the neighborhood except for the center cell $\mathbf n$, introduces the inter-cell coupling with $D$ being a rate constant, or diffusion constant. The third terms $\eta$ are the discrete counterparts of the thermal noises $\sigma$ in the TDGL; they are characterized by the following statistical properties
\begin{eqnarray}
	\label{eq:noise}
	\langle \eta_{j, \mathbf n}^t \rangle &=& 0, \nonumber \\
	\langle \eta_{i, \mathbf n}^t \, \eta_{j, \mathbf n'}^{t'} \rangle &=& 2 \Gamma_j \Delta t \, k_B T \,\delta_{ij}\, \delta_{\mathbf n, \mathbf n'} \, \delta_{t, t'}. 
\end{eqnarray}
For cell dynamics on a square grid, the neighborhood average $\langle\langle * \rangle\rangle$ is given in Ref.~\cite{oono88,puri88}. 
\begin{eqnarray}
	\langle\langle \phi \rangle\rangle = \frac{1}{6}\sum_{\rm NN} \phi + \frac{1}{12} \sum_{\rm NNN} \phi
\end{eqnarray}
where NN and NNN denote nearest-neighbor and second-nearest-neighbor sites, respectively. The operation $ \left( \langle \langle \phi \rangle \rangle - \phi  \right)$ is essentially the isotropically discretized Laplacian $\nabla^2\phi$. The effective diffusion coefficient $D$ can be related to lattice parameters as $D_j = {3 \Gamma_j  \Delta t}/{a^2}$, where $a$ is the lattice constant of the grid and $\Delta t$ is the time step. More generally, since both $\Gamma$, $\Delta t$ and $a$ are free parameters of the model, one can equally treat $D$ as a fitting parameter of the dynamical model.

The map function $f[*]$ is required to have two hyperbolic sinks and one hyperbolic source~\cite{oono87,teixeira97,sevink15}. Several maps have been proposed, without much effect on the simulation results. For a single order parameter TDGL, a popular choice is the tanh-map~\cite{oono88,chakrabarti92}: $f[\psi] = A \tanh(\psi)$, where parameter $A$ can be matched for particular systems of interest. Subsequently, several authors used a map function directly obtained from the free energy $\mathcal{F}$ and found that the cell dynamics method is still amenable for a realistic map function numerically~\cite{qi96,ren01}. In particular, one crucial advantage of using the derivative of the original free-energy for the map is to  include the effect of asymmetry of the two competing phases in e.g. a first-order phase transition~\cite{iwamatsu05,iwamatsu08}. It was later shown that such choice for cell dynamics can reproduce the essential features of the phase transformation kinetics even though the method is not guaranteed to be an accurate approximation of the original TDGL~\cite{teixeira97,sevink15}.
In our work, we adopt this approach in order to incorporate the $\lambda$-term into the cell dynamics simulations. The two mapping functions in our case are 
\begin{subequations}
\begin{eqnarray}
	f_1[\phi, \psi] &=& A_1 \phi - B_1 \phi^3 + C_1 \psi^2,  \\
	f_2[\phi, \psi] &=& A_2\psi - B_2\psi^3 + 2 C_2 \phi \psi, 
\end{eqnarray}
\end{subequations}
and the $A$, $B$, and $C$ coefficients are given by  
\begin{eqnarray}
	A_j = 1 + \Gamma_j \Delta t \,\tau_j, \quad B_j = \Gamma_j \Delta t \,g_j, \quad C_j = \Gamma_j \Delta t \lambda. \quad 
\end{eqnarray}
From these expressions, one can see that $\Gamma_j \Delta t$ is the effective time step. Specifically, here we use $ \Delta t = 0.01$, and set $\Gamma_1 = \Gamma_2 = 1$. For simplicity, we focus on the symmetric case $\tau_1 = \tau_2 \propto (T - T_c)$, and use parameters $g_1 = g_2 = 1$ and $D_1 = D_2 = 0.12$ in most of the simulations discussed in this paper.

\section{Phase ordering dynamics}
\label{sec:coarsening}

We first apply the cell dynamics method to study the coarsening dynamics of the coupled order parameters when the system is thermally quenched to the ordered phase. An initial random configuration is first annealed at a high temperature for several time steps. The reduced temperature is then abruptly decreased to a value  deep in the ordered phase. Since thermal fluctuations at such low temperatures have negligible effect on the phase-ordering process, they are neglected in our simulations below.  Fig.~\ref{fig:coarsening} shows the evolution snapshots of the charge and magnetic order parameters at different simulation times after the quench. We find that a well-developed charge order appears at $ t \gtrsim 5$ or roughly after 500 time-steps, while the magnetic order also exhibits well defined regions separated by domain-walls. Since the spin order parameter $\psi$ vanishes at the center of magnetic domain walls, the linear-quadratic coupling $\lambda \phi \psi^2$ results in a reduced charge order, or enhanced metallicity. Consequently, these magnetic domain walls manifest themselves as quasi-metallic strings in snapshots of the charge-order, consistent with the experimental observation~\cite{post18}. The evolution of the two order parameters are strongly correlated at the late stage: as the magnetic domain increases in size, the metallic strings of the charge-order become more sparse with time.

\begin{figure}[t]
\includegraphics[width=0.99\columnwidth]{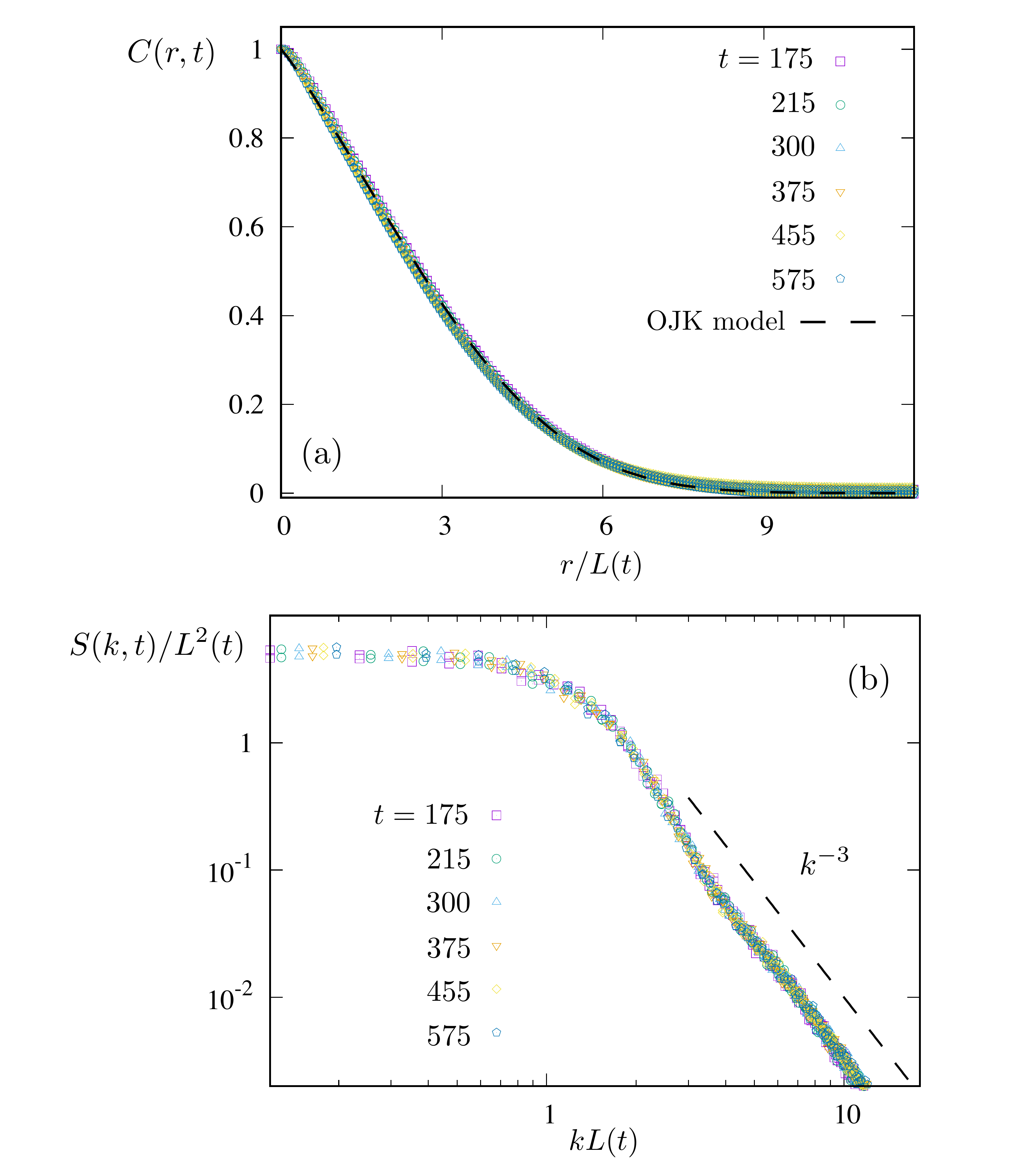}
\caption{(Color online)  
\label{fig:scaling} Scaling plot of correlation function $C(r, t)$ and structure factor $S(k, t)$ of the magnetic order parameter field $\psi(\mathbf r, t)$. (a) Plot of spherically averaged $C(r, t)$ versus scaled distance $r / L(t)$ at different simulation times. The dashed line corresponds to the analytical OJK function. (b) Log-log plot of the spherically averaged scaled structure factor $S(k,t)/L^2$ versus scaled momentum $k L(t)$. The dashed line indicates the $k^{-3}$ Porod's law behavior in 2D. In both plots, a power-law time dependence is used for the length scale $L(t) \sim t^{1/z}$ with an exponent $z = 2$.
}
\end{figure}

In order to study the morphology of the domain growth at the late stage, we compute the equal-time correlation function of the magnetic order parameter
\begin{eqnarray}
	\label{eq:corr}
	C(\mathbf r, t) = \langle \psi(\mathbf r_0 + \mathbf r, t) \psi(\mathbf r_0, t) \rangle - \langle \psi(\mathbf r_0, t)\rangle^2
\end{eqnarray}
Here the angular brackets denote average over the reference point $\mathbf r_0$ and over independent runs with different initial conditions. The Fourier transform of the correlation function corresponds to the structure factor which can be measured in scattering experiments,
\begin{eqnarray}
	S(\mathbf k, t) = \int C(\mathbf r, t) e^{i \mathbf k \cdot \mathbf r} \, d\mathbf r.
\end{eqnarray}
The correlation function and structure factor of a non-conserved order parameter is expected to follow certain scaling behaviors during coarsening. 
Fig.~\ref{fig:scaling}(a) shows the normalized $C(r, t)$ versus scaled distance $r / L(t)$ at different simulation times during the late stage of the phase ordering. Assuming a power-law increase for the characteristic length $L(t) \sim t^{1/z}$ with an exponent $z = 2$, we find that the scaled correlation functions at different times collapse into a universal function $g(x)$, 
\begin{eqnarray}
	C(\mathbf r, t) = g\!\left(\frac{|\mathbf r|}{L(t)} \right)
\end{eqnarray}
which indicates a scaling behavior for the growth of magnetic domains. We also compare this universal function with the analytical correlation obtained by Ohta, Jasnow, and Kawasaki (OJK) for the motion of random interfaces~\cite{ohta82}
\begin{eqnarray}
	C_{\rm OJK}(r, t) = \frac{2}{\pi} \sin^{-1}\!\left[ \exp\left(-\frac{r^2}{8 D t} \right) \right]
\end{eqnarray}
where $D$ is an effective diffusion constant. The above correlation function also implies a characteristic length scale $L(t) \sim \sqrt{8 D t} \sim t^{1/2}$, consistent with the well-known Allen-Cahn law for the domain growth of non-conserved order parameter~\cite{bray02,puri09}. As shown in Fig.~\ref{fig:scaling}(a), the collapsed data points agree reasonably well with the OJK correlation function.

\begin{figure}
\includegraphics[width=0.99\columnwidth]{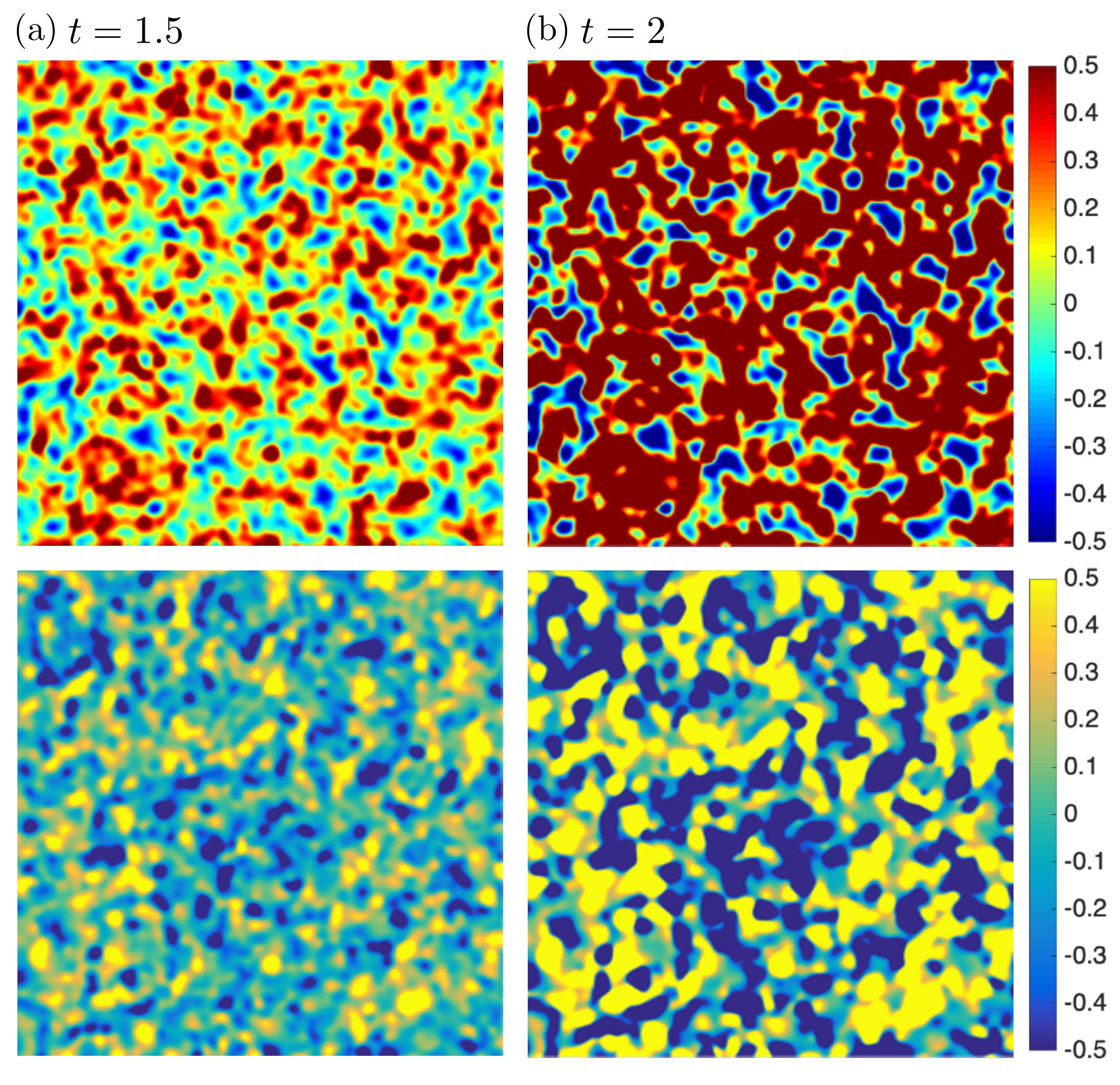}
\caption{(Color online)  
\label{fig:coarsening2} Snapshots of early stage of coarsening process when the system is quenched to a reduced temperature $\tau^* = -1.9$. The panels show a $300\times300$ area of charge $\phi$ (top) and spin $\psi$ (bottom) order parameters obtained from cell dynamics simulations on a $2000\times 2000$ square-lattice grid. The charge-spin coupling is $\lambda = 2$.  The time-step used in the simulation is~$\Delta t = 0.01$. In contrast to snapshots in Fig.~\ref{fig:coarsening}, here charge domains of both positive ($\phi > 0$) and negative ($\phi < 0$) values can be seen in the early stage of the phase ordering.   
}
\end{figure}

Similar scaling behavior is also obtained for the structure factor. Fig.~\ref{fig:scaling}(b) shows the log-log plot of $S(k, t) / L^2(t)$ versus scaled momentum $k L(t)$ at different times with a power-law $L(t) \sim t^{1/2}$. Again, the nice data-point collapsing confirms the scaling behavior and the existence of universal scaling function $\tilde{g}$
\begin{eqnarray}
	S(\mathbf k, t) = L^D(t) \,\tilde{g}\bigl(|\mathbf k| L(t) \bigr).
\end{eqnarray}
where $D = 2$ for two dimensions. 
Moreover, in the large $k$ limit, the structure factor is found to decay as $S(k, t) \sim k^{-3}$, consistent with the Porod's law~\cite{porod82}. This special power-law at large $k$ results from the sharp interfaces formed between the two degenerate magnetically ordered states. 

Our analysis above shows that the coarsening dynamics of the magnetic order $\psi$, which preserves a global $Z_2$ symmetry,  obeys the Allen-Cahn theory of domain growth for non-conserved order parameter~\cite{bray02,puri09}, such as the Ising model. The presence of the charge-order $\phi$ does not qualitatively affect the power-law scaling of the coarsening process. Similar results have also been observed for the non-conserved order parameter in the model-C kinetics when the system is quenched into the ordered phase~\cite{kockelkoren02,das17}. However, a different power-law $L\sim t^{1/3}$ domain-growth, characteristic of a conserved field, was observed for quenches into the coexistence regime in model~C~\cite{kockelkoren02}.

\begin{figure}
\includegraphics[width=0.95\columnwidth]{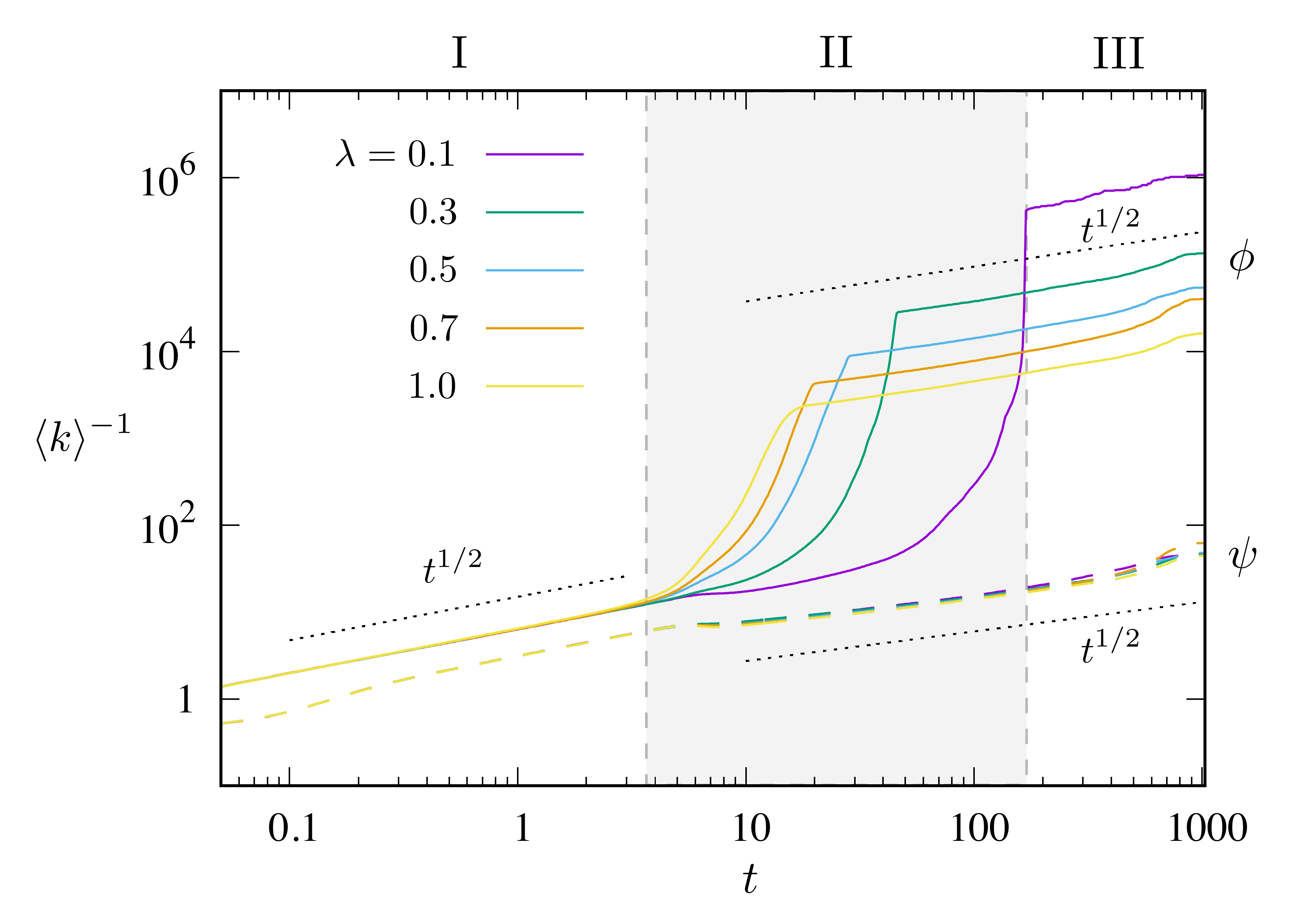}
\caption{(Color online)  
\label{fig:Lt} Characteristic length $L(t) \sim \langle k \rangle^{-1}$ calculated from inverse of the average wave vector versus time for different values of the coupling $\lambda$. The solid and dashed lines correspond to the charge $\phi$ and magnetic $\psi$ order parameters, respectively.   The dotted lines represent the Allen-Cahn growth law. The time-step used in the simulations is $\Delta t = 0.01$.
}
\end{figure}

While the charge-order essentially becomes slave to the dynamics of the magnetic order parameter at late stage of the phase ordering, detailed examination of the snapshots, for example Fig.~\ref{fig:coarsening}(a) and Fig.~\ref{fig:coarsening2}, shows that the initial evolution of the two order parameters are decoupled from each other even with a large nonzero charge-spin coupling. To further study this dynamical decoupling, we compute time-dependent characteristic length scales of the two order-parameter fields based on the spherically averaged wave vector defined as
\begin{eqnarray}
	\langle k \rangle = \frac{ \int |\mathbf k| \, S(\mathbf k, t) d\mathbf k}{ \int S(\mathbf k, t) d\mathbf k}.
\end{eqnarray} 
Here $S(\mathbf k, t)$ is the structure factor obtained from the charge or the spin configurations. We then define length scales of the two fields as $L_\phi(t) \sim 1/\langle k \rangle_\phi$ and $L_\psi(t) \sim 1/\langle k \rangle_\psi$, respectively.  Fig.~\ref{fig:Lt} shows these two length scales versus time from simulations with different values of $\lambda$. Interestingly, as shown in regime I of the plot, both order parameters exhibit a seemingly power-law domain growth immediately after quench. In this first stage of phase ordering, the evolutions of the two scalar fields are not strongly correlated; see Fig.~\ref{fig:coarsening2}. In particular, the dynamics of the charge-order seems to preserve a $Z_2$ symmetry, which means both positive and negative $\phi$-domains (red and blue regions, respectively, in top panels of Fig.~\ref{fig:coarsening2}) are expanding. As shown in Fig.~\ref{fig:Lt}, the domain growth in regime I follows the  Allen-Cahn law expected for non-conserved scalar field.

The charge-spin coupling starts to exert its effect in the transient dynamical regime II, where domains with negative $\phi$, shown as blue regions in Fig.~\ref{fig:coarsening2},  start to disappear. The duration of this re-alignment period is controlled by the coupling strength $\lambda$. The growth of the magnetic domains also slightly slows down during this period. This can be attributed to the fact that nonzero magnetic order parameter only develops inside the positive-$\phi$ regions. Consequently, the expansion of the magnetic domains has to wait until the disappearance of the surrounding energetically unfavorable negative-$\phi$ puddles. The second dynamical regime ends when all the negative-$\phi$ domains disappear, and the overall phase-ordering is now controlled by the coarsening dynamics of the magnetic order parameter, which is shown to follow the $L \sim t^{1/2}$ Allen-Cahn law.

\section{nucleation and growth}
\label{sec:nucleation}

\begin{figure*}
\includegraphics[width=1.99\columnwidth]{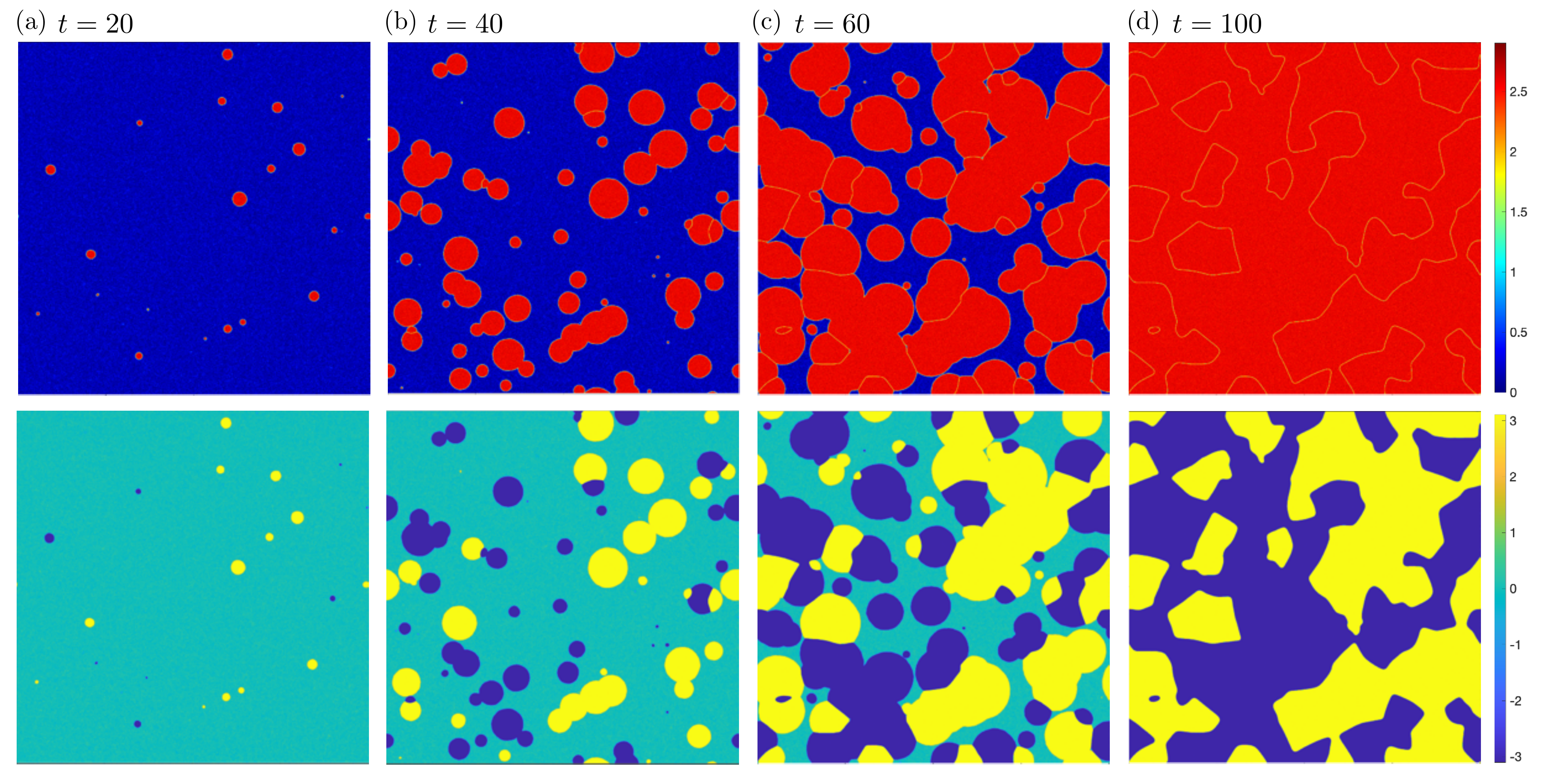}
\caption{(Color online)  
\label{fig:nucleation1} Snapshots of the nucleation and growth when the system is quenched from the high-temperature paramagnetic phase to a reduced temperature $\tau^* = 0.565$ and $ \lambda = 2$ in the coexistence regime, obtained from cell dynamics simulations on a $2000\times 2000$ square grid.  The top and bottom panels show configurations of the charge $\phi$ and spin $\psi$ order parameters, respectively. The scalar field $\phi$ also serves as the order-parameter for metal-insulator transition, with nonzero $\phi$ (red areas) correspond to insulating regions. The time-step used in the simulations is $\Delta t = 0.01$. 
}
\end{figure*}

We next study the nucleation and growth dynamics when the system is quenched to the coexistence regime of the first-order transition from either the high-temperature paramagnetic phase, or the low-temperature ordered state. Although we focus on quench simulations for convenience, our results also have implications to the adiabatic annealing or warming protocols used in the experiments~\cite{post18}. For example, as the temperature is slowly decreased during an annealing process, the system remains in the quasi-equilibrium disordered state, even at temperatures where the high-$T$ paramagnetic phase becomes unstable. Transition to the low-$T$ phase occurs only when thermal fluctuations are strong enough to overcome the free-energy barrier and initiate the nucleation process. The temperature $T^*$ at which the phase transformation takes place depends on the annealing rate and the strength of thermal noise. For slow enough annealing, this is similar to quenching the system to a particular temperature $T^*$, from the numerical point of view.

In contrast to phase-ordering simulations discussed in the previous section, thermal noise plays an important role in the cell-dynamics simulation of nucleation and growth.  As indicated in Eq.~(\ref{eq:noise}), thermal fluctuations are modeled by a spatially un-correlated white noise with zero mean and a variance proportional to temperature, the time-step $\Delta t$, and the effective relaxation coefficient~$\Gamma$. Since the width of the coexistence window in general is much smaller compared with the critical temperature $T_c$, we can thus neglect the temperature dependence of the thermal noise and approximate the temperature in Eq.~(\ref{eq:noise}) by $T_c$. On the other hand, as our model is defined in terms of the reduced temperatures $\tau_j$, the critical temperature $T_c$ is a free parameter, which means we are free to choose the variance of the thermal noise in cell dynamics. In our quench simulations discussed below, we use variances ${\rm Var}(\phi) = {\rm Var}(\psi) = 0.005$. To incorporate thermal fluctuations in our simulation, at each time step we first perform the deterministic evolution of the cell dynamics, i.e. the first two terms in Eq.~(\ref{eq:cds_eq}). Then a random number sampled from a normal distribution with zero mean and fixed variance specified above is added to the two scalar variables at every site.

\subsection{Metal-to-insulator transition from the paramagnetic phase}

We first consider cell dynamics simulations for nucleation from the high-temperature paramagnetic phase. Starting from a random configuration, the system is first equilibrated at a high enough reduced temperature for thousands of time steps. Then it is  cooled to a temperature $\tau^*$ that lies in the coexistence window in Fig.~\ref{fig:phase}(b). Snapshots of nucleation and growth obtained from cell dynamics simulations with $\tau^* = 0.565$ and $\lambda = 2$ are shown in Fig.~\ref{fig:nucleation1}. The initial state with both order-parameters being zero is a meta-stable state at this temperature, and is unstable against decaying into the ordered phase. As shown in Fig.~\ref{fig:nucleation1}(a), nearly circular nuclei of the new phase appear spontaneously and (spatially) randomly from thermal fluctuations. It is worth noting that magnetic domains of both positive and negative $\psi$, yellow and blue domains in Fig.~\ref{fig:nucleation1}(a), are independently nucleated due to thermal noise. Moreover, these droplets of ordered phase remain more or less circular during their growth. As these droplets further expand, they start to collide with each other and merge to form larger domains, until eventually the whole system is transformed into the stable ordered phase.

\begin{figure}[t]
\includegraphics[width=0.99\columnwidth]{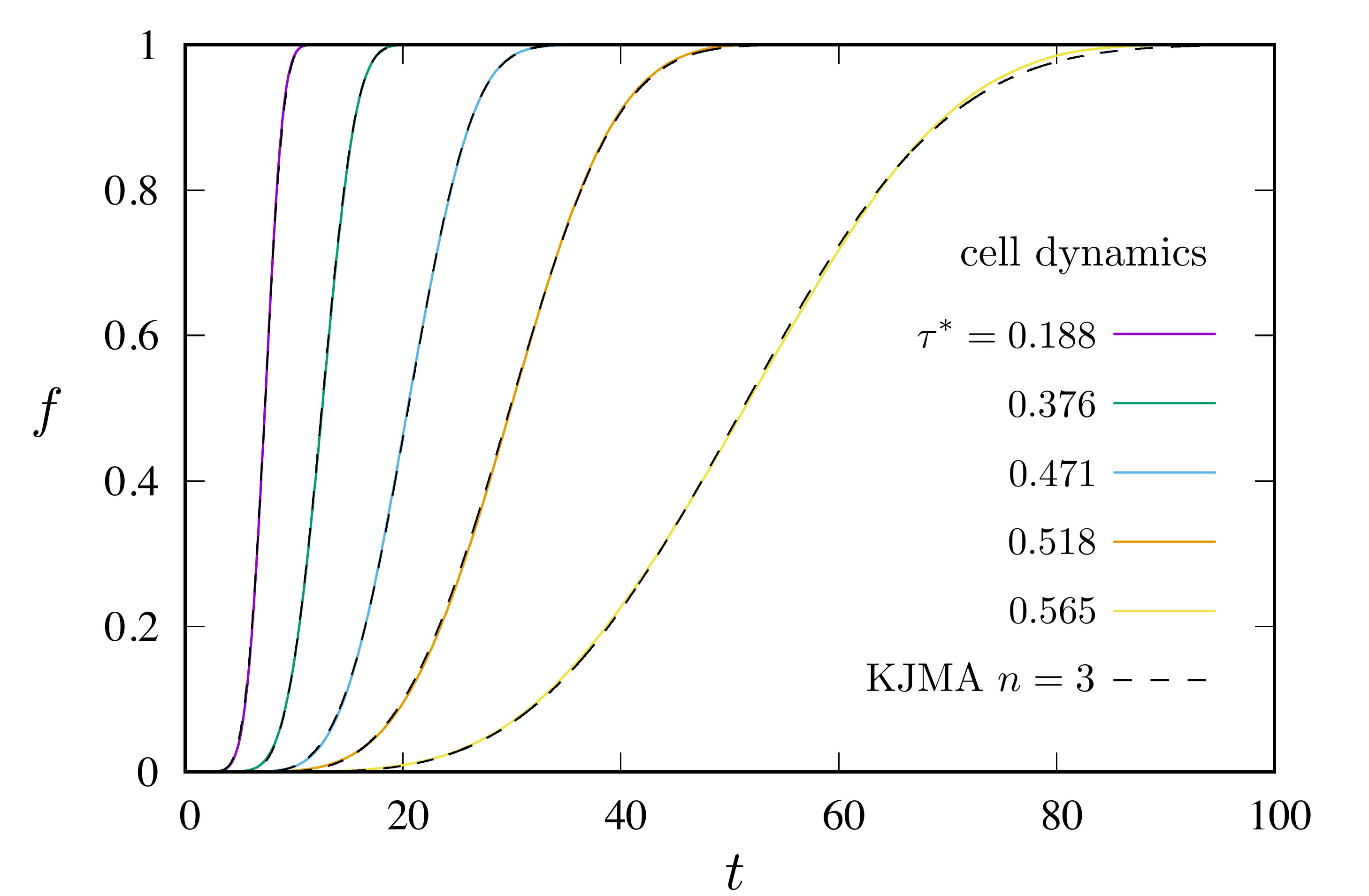}
\caption{(Color online)  
\label{fig:KJMA} Time dependence of the relative area $f$ that is covered by the transformed new phase for quench into varying reduced temperatures $\tau^*$ in the coexistence window of the first-order transition.  The charge-spin coupling is set to $\lambda = 2$ in our units. Dashed lines represent modified KJMA formula Eq.~(\ref{eq:KJMA}) with an exponent $n = 3$. 
}
\end{figure}

A quantitative analysis of the transition dynamics is the time dependence of the volume fraction $f$ of the transformed phase, which is shown in Fig.~\ref{fig:KJMA} for different temperatures in the coexistence window. The overall transformation rate is faster with decreasing quench temperature $\tau^*$, consistent with the reduced energy barrier separating the meta-stable high-$T$ paramagnetic phase and the stable low-$T$ ordered state. The kinetics of such nucleation and growth phenomena is often described using the Kolmogorov-Johnson-Mehl-Avrami (KJMA) theory~\cite{kolmogorov37,avrami39,avrami40,johnson39}, which relates the volume fraction $f$ to the so-called extended volume fraction~$\Omega(t)$
\begin{eqnarray}
	f(t) = 1 - \exp[- \Omega(t) ].
\end{eqnarray}
Here $\Omega$ can be viewed as the relative volume or area of the transformed phase in the absence of other nuclei. For $D$-dimensional nucleation, It is given by the formula $\Omega(t) = \int_0^t I(t')\, R^D(t; t') \, dt'$, where $I(t)$ is the nucleation rate and $R(t; t')$ is the radius of nucleus at time $t$ that was nucleated at $t'$. Assuming a constant nucleation rate $I(t) = I_s$ as well as constant a size-independent domain growth, which implies a linear dependence of the nuclei radius $R(t; t') = v_s (t - t')$, where $v_s$ is the interfacial velocity, one obtains the well known KJMA formula
\begin{eqnarray}
	\label{eq:KJMA}
	f(t) = 1 - \exp\left[-K (t - t_{\rm inc})^n\right].
\end{eqnarray}
Here $n = D + 1$ is the so-called Avrami exponent, the coefficient $K \propto I_s v_s^D$, and we have included a finite incubation time $t_{\rm inc}$ for nucleation. In two dimensions, we have $K = \frac{\pi}{3} I_s v_s^2$ and $n = 3$.  The nucleation rate is related to the free-energy barrier $W_*$: 
\begin{eqnarray}
	\label{eq:I_s}
	I_s \propto \exp(-W_*/k_B T). 
\end{eqnarray}
In order to use the KJMA formula to describe our simulation results, we find that it is crucial to include a finite~$t_{\rm inc}$. There are two major contributions to the incubation time
\begin{eqnarray}
	\label{eq:t_inc}
	t_{\rm inc} = t_* \ln\left(\frac{t}{t_*} \frac{W_*}{k_B T} \right) - t_*,
\end{eqnarray}
where $t_*$ is a time-scale related to the finite radius $R_*$ of critical nucleus. Taking into account this initial value, the time dependence of the nuclei radius needs to be modified as $R(t; t') \approx v_s (t - t') + R_*$. This is equivalent to shifting the origin of the time scale in the past direction by $t_* = R_*/ v_s$, corresponding to the second term in Eq.~(\ref{eq:t_inc}).  The first term, on the other hand, comes from a size-dependent growth rate related to the transient nucleation effects~\cite{shneidman93}. 
For cooling into a meta-stable state with large energy barrier $W_*$, a critical nucleus is rarely formed and the time necessary to reach the steady state would be long. Consequently, the first term in Eq.~(\ref{eq:t_inc}) dominates, giving rise to a positive $t_{\rm inc}$, which is the case in most of our simulations.

\begin{figure}[t]
\includegraphics[width=0.97\columnwidth]{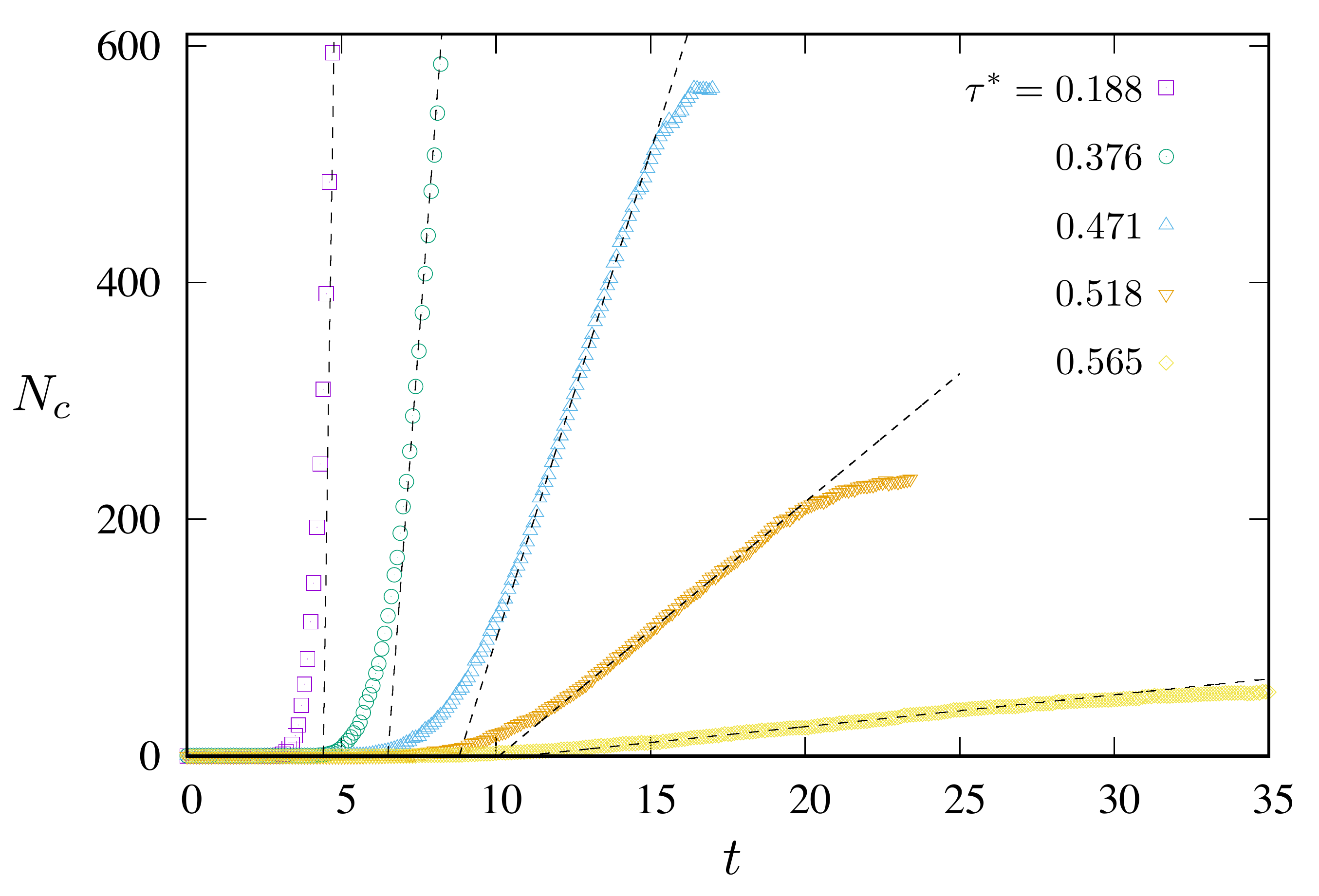}
\caption{(Color online)  
\label{fig:cls_num} Time dependence of the number of nuclei or clusters of transformed phase for quench into various reduced temperatures. The dashed lines correspond to a linear fit according to Eq.~(\ref{eq:linear_t}).
}
\end{figure}

\begin{figure}[t]
\includegraphics[width=0.97\columnwidth]{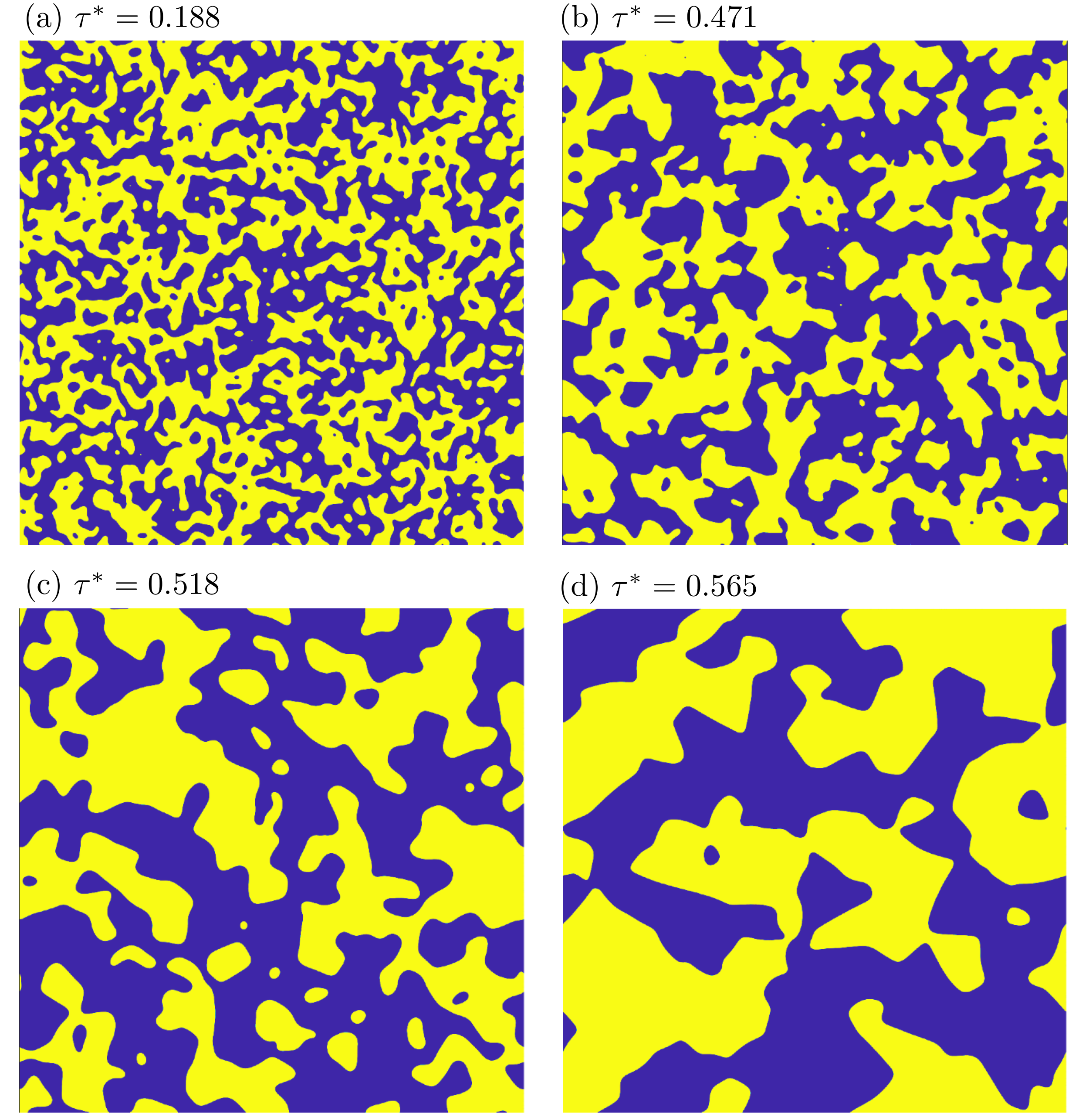}
\caption{(Color online)  
\label{fig:mag-dm} Snapshots of magnetic order parameter field $\psi(\mathbf r)$, obtained from simulations on a $2000\times 2000$ square lattice, at the end of phase transformation in Fig.~\ref{fig:KJMA} for varying reduced temperatures $\tau^*$.  
}
\end{figure}


As shown in Fig.~\ref{fig:KJMA}, the transformation curves $f(t)$ obtained from our simulations can be well described by the modified KJMA formula~(\ref{eq:KJMA}) with $n = 3$, taking into account both the transient nucleation effects and finite size of critical nuclei. Importantly, this agreement indicates that the transformation into the coupled charge- and spin-ordered state is characterized by a constant nucleation rate and interfacial velocity. This constant nucleation regime at small $t$ also manifests itself in the linear increase of the number of nucleating clusters with time:
\begin{eqnarray}
	\label{eq:linear_t}
	N_c(t) \approx I_s\, (t - t_0),
\end{eqnarray}
where $I_s$ is the steady-state nucleation rate, and $t_0$ is an empirical constant called the induction time~\cite{papon06}. Such linear dependence is explicitly verified in our simulations, as shown in Fig.~\ref{fig:cls_num}. 

The metal-insulator transition ends when insulator domains with $\phi > 0$, i.e. red regions in Fig.~\ref{fig:nucleation1}, occupy the whole sample. On the other hand, due to the $Z_2$ symmetry of the magnetic order, multiple magnetic domains emerge at the end of the phase transformation, as illustrated in Fig.~\ref{fig:nucleation1}(d). Upon lowering the temperature, the system will then undergo a coarsening process discussed in Sec.~\ref{sec:coarsening}. The size of magnetic domains at the end of phase transformation depends on the quench temperature. For example, Fig.~\ref{fig:mag-dm} shows snapshots of magnetic order-parameter field $\psi$ at four different reduced temperatures  $\tau^*$ when transition to the $\phi \sim 0$ insulating phase is completed. As can be seen from these figures, the domain size increases with higher quench temperatures. 

One can estimate the typical linear size $\ell_m$ of these magnetic domains as follows. A domain wall forms when expanding nuclei of opposite magnetic order parameter collide with each other. Consequently, the typical domain size is $\ell_m \sim v_s \, \tau_c$, where $v_s$ is the interfacial growth velocity, and $\tau_c$ is the average growth time before collision. During this time scale, the number density of nuclei is expected to be $n_c \sim I_s \, \tau_c$. This corresponds to a length scale $1/\sqrt{n_c}$ which is basically the average distance between nuclei. By identifying this length scale with $\ell_m$, we have $1/\sqrt{I_s \tau_c} \sim v_s \tau_c$, or $\tau_c \sim (1/I_s v_s^2)^{1/3}$. Substitute this estimate of $\tau_c$ into the expression for $\ell_m$, we obtain
\begin{eqnarray}
	\label{eq:l_m}
	\ell_m \sim \left(\frac{v_s}{I_s} \right)^{1/3} \sim v_s^{1/3} \, \exp\left(\frac{W_*}{3 k_B T} \right).
\end{eqnarray}
Here we have used Eq.~(\ref{eq:I_s}) for the dependence of nucleation rate $I_s$ on the energy barrier. This result shows that the typical domain size increases exponentially with the barrier $W_*$. 

\begin{figure}[t]
\includegraphics[width=0.97\columnwidth]{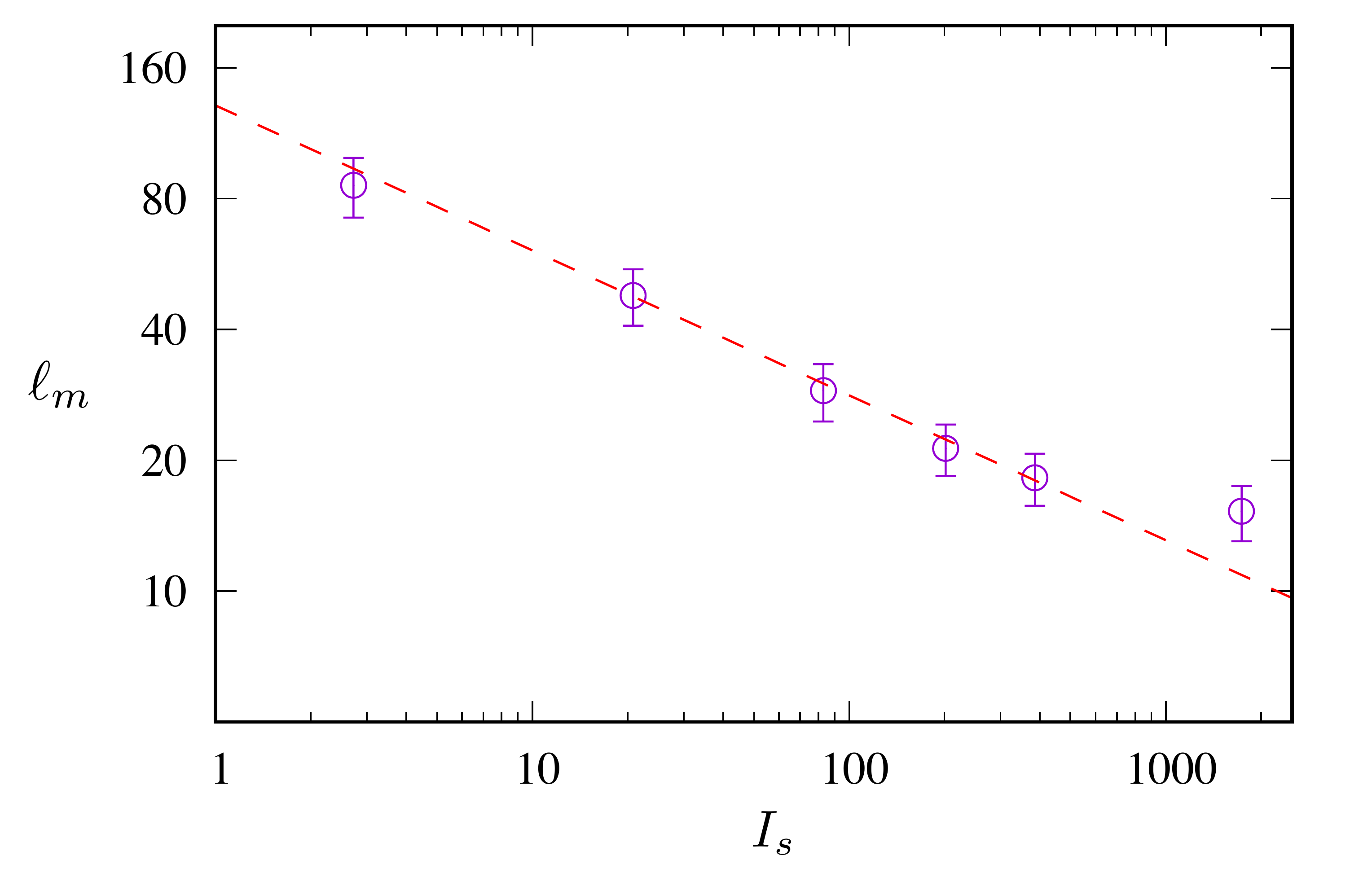}
\caption{(Color online)  
\label{fig:dm-size} Log-log plot of domain size $\ell_m$ versus nucleation rate~$I_s$ obtained from cell-dynamics simulations on a $2000\times 2000$ square lattice. The dashed line corresponds to the power-law $\ell_m \sim I_s^{-1/3}$ dependence indicated in Eq.~(\ref{eq:l_m}). 
}
\end{figure}

To verify the power-law dependence between the domain-size and nucleation rate, we numerically compute~$\ell_m$ from the correlation function Eq.~(\ref{eq:corr}) of magnetic order-parameter $\psi$ at the end of phase transformation, i.e. $\ell_m \sim \sum_{\mathbf r} |\mathbf r| C(\mathbf r) / \sum_{\mathbf r} C(\mathbf r)$. On the other hand, using Eq.~(\ref{eq:linear_t}), we estimate the nucleation rate from the slope $I_s \sim dN_c / dt$ of the linear segment of the $N_c(t)$ curves shown in Fig.~\ref{fig:cls_num}. Comparing these two quantities at the same quench temperature, our results summarized in Fig.~\ref{fig:dm-size} show a nice agreement with the $\ell_m \sim I_s^{-1/3}$ power-law relation, except at very large nucleation rate with high density of nuclei.

\begin{figure*}[t]
\includegraphics[width=1.99\columnwidth]{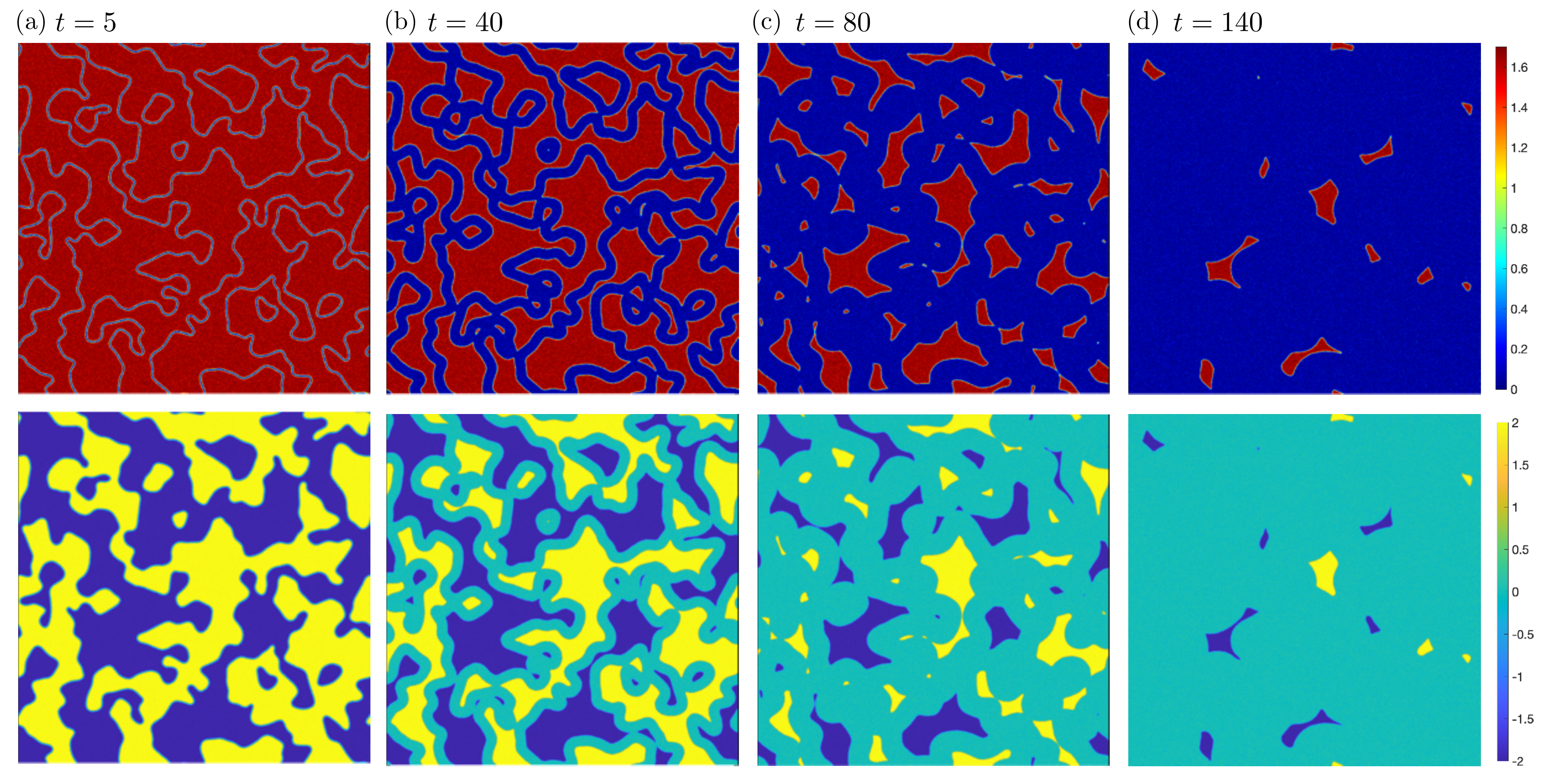}
\caption{(Color online)  
\label{fig:nucleation2} Snapshots of the nucleation and growth when the system is quenched from the low-temperature ordered phase to a reduced temperature $\tau^* = 2.353$ and $ \lambda = 2$ in the coexistence regime, obtained from cell dynamics simulations on a $2000\times 2000$ square grid.  The top and bottom panels show configurations of the charge $\phi$ and spin $\psi$ order parameters, respectively. The scalar field $\phi$ also serves as the order-parameter for metal-insulator transition, with nonzero $\phi$ (red areas) correspond to insulating regions. The time-step used in the simulations is $\Delta t = 0.01$. 
}
\end{figure*}

\subsection{Insulator-to-metal transition with pre-existing magnetic domain walls}
\label{sec:dw-nucleation}

Next we consider the transformation from the low-temperature ordered states to the high-$T$ paramagnetic phase. Interestingly, we find that for a wide range of parameters the Landau theory given in Eq.~(\ref{eq:landau}) admits a very narrow temperature regime (near the high end of the coexistence window) where the ordered state is meta-stable. We have considered two different phase transformation scenarios, depending on whether there are pre-existing magnetic domain walls in the initial state.  In the absence of multiple magnetic domains, thermal noise is required to initiate the nucleation of high-$T$ paramagnetic phase. Our simulations find that the transformation kinetics can again be well described by the two-dimensional KJMA formula~(\ref{eq:KJMA}) with $n = 3$ when the transient nucleation effects are properly included.

\begin{figure}[b]
\includegraphics[width=0.99\columnwidth]{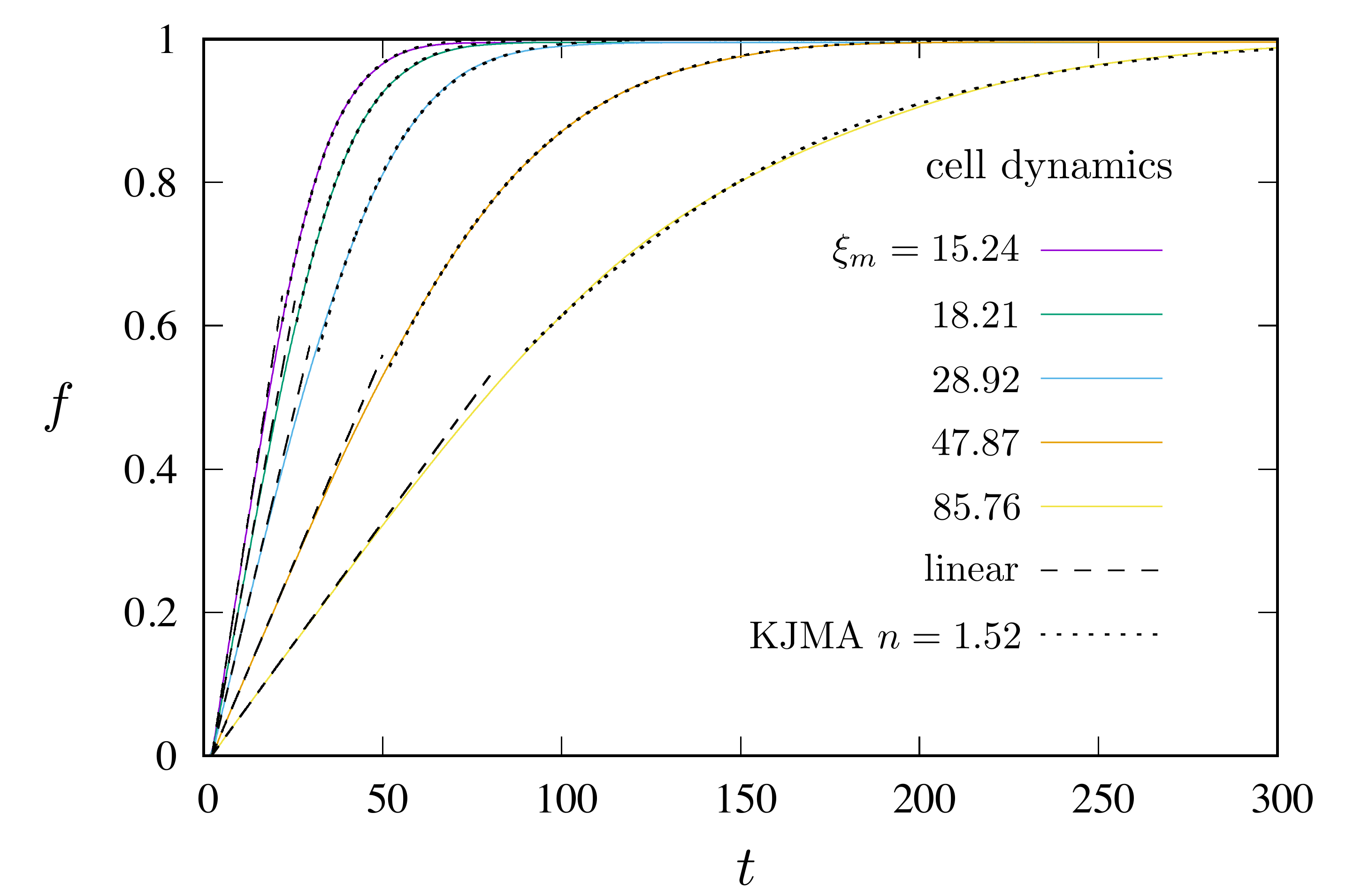}
\caption{(Color online)  
\label{fig:f_metal} Time dependence of the transformed volume fraction for quench from the low-$T$ ordered state to the a temperature $\tau^* = 2.353$ in the coexistence window. The charge-spin coupling is set to $\lambda = 2$ in our units. Different curves here correspond to different magnetic correlation length $\xi_{m}$ in the initial state.  Dashed lines represent linear increase at small $t$, while dotted lines represent modified KJMA formula Eq.~(\ref{eq:KJMA}) with an exponent $n = 1.52$.   
}
\end{figure}

The presence of qausi-metallic magnetic domain walls in the initial states dramatically changes the phase transformation dynamics. Essentially, these domain walls serve as efficient seed for nucleation of the metallic domains. Consequently, thermal fluctuations do not play a crucial role in this scenario. Fig.~\ref{fig:nucleation2} shows snapshots of the two order-parameter fields during the insulator-to-metal transition when an initial ordered state becomes meta-stable under a temperature quench. We have verified that nucleation and growth of the metallic phase take place even without  thermal noises. Instead, as shown in Fig.~\ref{fig:nucleation2}(a) and (b), the new phase materializes around the pre-existing quasi-metallic strings or magnetic domain-walls. 

Fig.~\ref{fig:f_metal} shows the volume fraction of the transformed metallic phase as a function of time after the quench. The different curves correspond to initial state with different magnetic correlation length $\xi_{m}$. Numerically, this length scale is computed from the magnetic correlation function defined in Eq.~(\ref{eq:corr}). The initial stage of the phase transformation can be well described by a linear function (dashed lines in Fig.~\ref{fig:f_metal})
\begin{eqnarray}
	\label{eq:linear_growth}
	f(t) \approx \alpha_s\, (t - t_d)
\end{eqnarray}
where $\alpha_s$ characterizes the growth rate and $t_d$ is called the decay time of the domain-wall to be discussed below. We find that while time scale $t_d$ is nearly independent of the initial magnetic correlation length, the growth rate~$\alpha_s$ is roughly inversely proportional to the initial~$\xi_m$.

\begin{figure}[t]
\includegraphics[width=0.99\columnwidth]{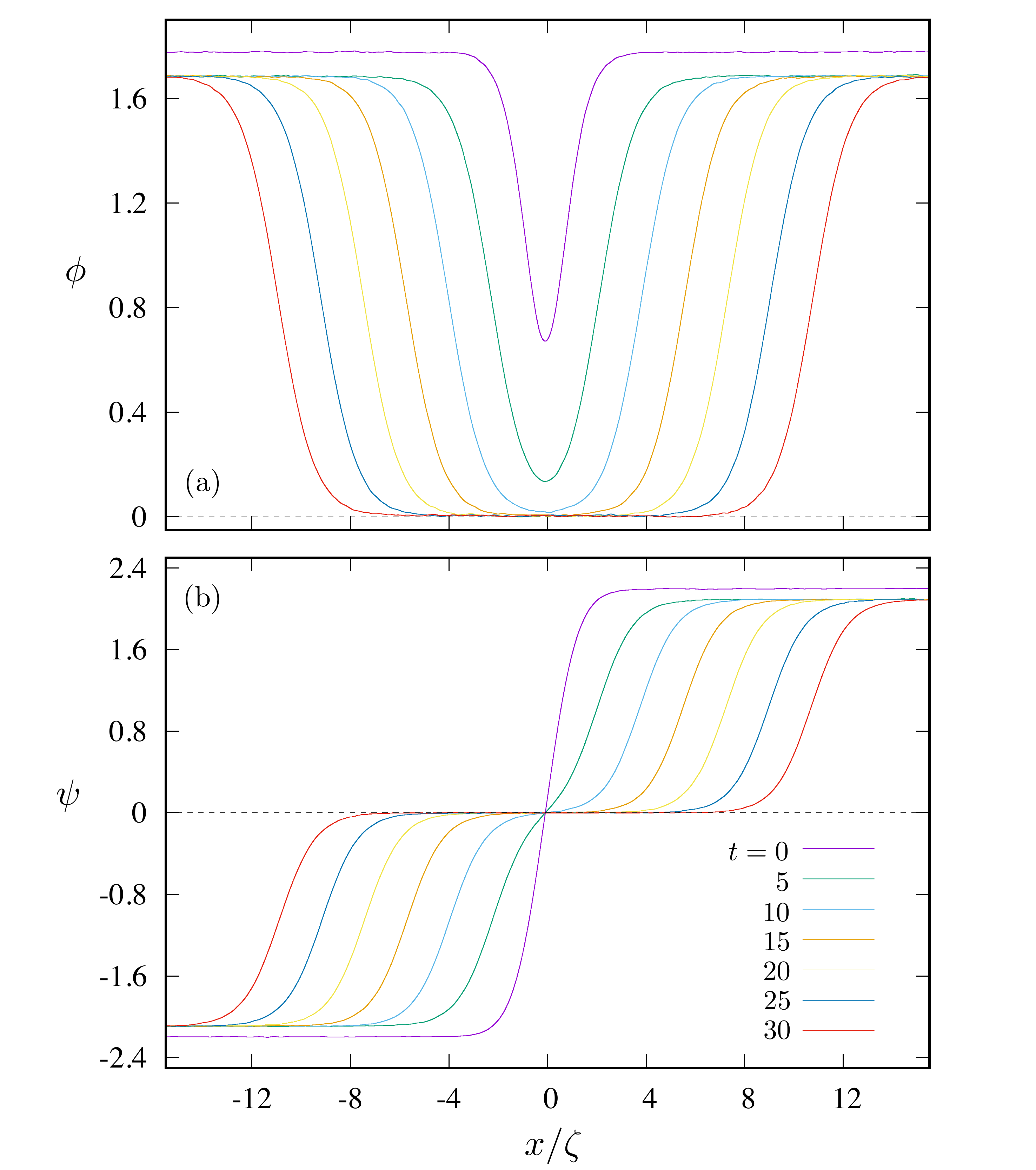}
\caption{(Color online)  
\label{fig:dw} Evolution of an initial magnetic domain-wall configuration under a thermal quench. $\zeta$ is a length scale characterizing the domain-wall width in the ordered state. 
}
\end{figure}

These interesting behaviors are in fact a direct consequence of the domain-wall induced nucleation mechanism. To see this, we numerically solve the one-dimensional coupled TDGL equation starting from an initial configuration corresponding to a  magnetic domain wall at the center, i.e. $\psi(x, t = 0) \to \pm \psi_{\rm eq}$ as $x \to \pm \infty$. The time evolution of the two scalar fields are shown in Fig.~\ref{fig:dw}. Due to the charge-spin coupling, the charge-order parameter $\phi(x)$ develops a dip at the center $x = 0$, giving rise to enhanced metallicity compared with the bulk of the domain. As the insulating state becomes unstable after the quench, this dip becomes deeper and eventually a metallic region with $\phi = 0$ develops at the center. This process corresponds to the initial nucleation ($t < t_d$) of the transformation curves in Fig.~\ref{fig:f_metal}. Crucially, the time scale $t_d$ of this initial nucleation process depends only on properties of the magnetic domain-wall, and is independent of the initial correlation length $\xi_m$ which determines the density of domain walls.

\begin{figure}
\includegraphics[width=0.99\columnwidth]{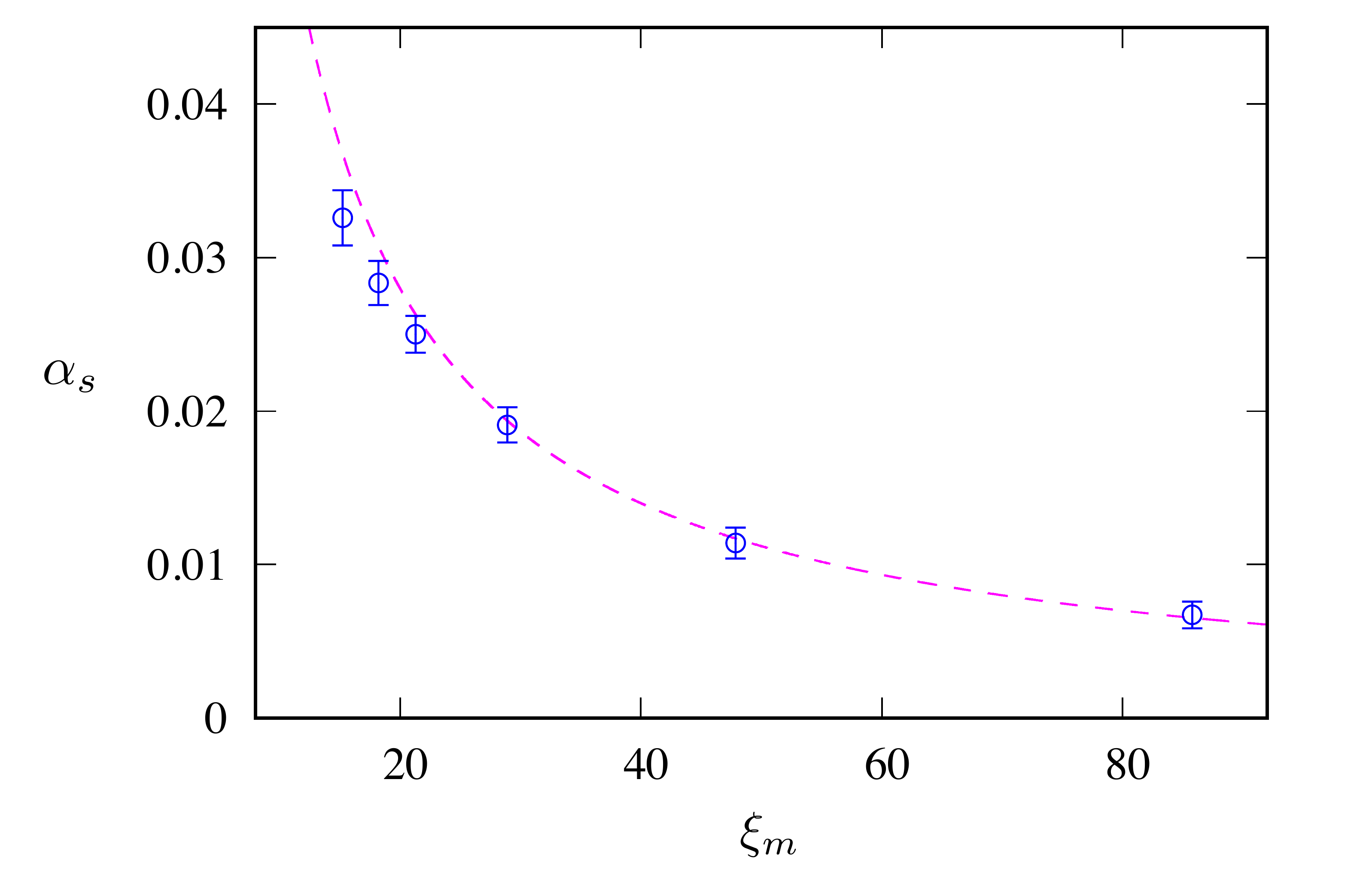}
\caption{(Color online)  
\label{fig:growth-rate} The linear growth rate $\alpha_s$ versus the initial correlation length $\xi_m$ of the magnetic order parameter.
}
\end{figure}

Next we consider the mechanism of the linear transformation rate. To facilitate the discussion, we use the terminology introduced in Ref.~\cite{post18} and introduce the insulator-metal boundary (IMB) which separates the metallic region with $\phi \approx 0$ from the insulator domain characterized by $\phi > 0$. As shown in Fig.~\ref{fig:dw}(b), the nucleation process can then be viewed as the decay of the magnetic domain-wall into two such IMB's. Once formed, these two IMB's then move away from each other with a constant velocity, giving rise to a steady growth of the metallic region sandwiched by these two interfaces. In this picture, the total area or ``volume" of the transformed metallic phase is well approximated by $f(t)\sim \mathcal{L}_{\rm dw} \times 2 v_{\rm IMB} \times (t - t_0)$, where $\mathcal{L}_{\rm dw}$ is the total length of domain walls in the initial state, $v_{\rm IMB}$ is the propagation velocity of the IMB. We can estimate the total length $\mathcal{L}_{\rm dw}$ as follows. For a given correlation length~$\xi_m$, the average size of a magnetic domain is roughly $\pi \xi_m^2$. The number of magnetic domains is thus  $N_m = A/(\pi \xi_m^2)$, where $A$ is the area of the system. The total length can then be estimated as the total circumference of these domains $\mathcal{L}_{\rm dw} \sim N_m \times 2 \pi \xi_m \sim A/\xi_m$. The transformation rate $\alpha_s = d f / dt$ is thus
\begin{eqnarray}
	\alpha_s \sim v_{\rm IMB} / \xi_m.
\end{eqnarray}
This inverse proportionality is verified in our numerical simulations, as shown in Fig.~\ref{fig:growth-rate}.

Finally, one can also understand the linear growth rate Eq.~(\ref{eq:linear_growth})  from the general theory of Avrami exponent $n = \alpha + D p$, where $D$ is spatial dimension, $p$ is related to growth mechanism, and the constant $\alpha$ depends on the nucleation rate. Specially, $p = 1$ for interface-controlled growth, while $p = 1/2$ for diffusion-dominated growth. And constant nucleation rate corresponds to $\alpha = 1$, while pre-existing nuclei is described by $\alpha = 0$. Since in our case, the pre-existing magnetic domain-walls serves as the seed of nucleation and the growth is controlled by the propagation of IMB, we have $\alpha = 0$ and $p = 1$, giving rise to an Avrami exponent $n = D$. The linear-growth rate thus indicates an effective spatial dimension $D = 1$, which is consistent with the fact that the 1D nature of domain-wall forces the phase transformation to take place only in the perpendicular direction. 
Interestingly, we find that the late stage of the phase transformation can be well approximated by the KJMA equation~(\ref{eq:KJMA}) with an exponent $n \approx 1.52$; see Fig.~\ref{fig:f_metal}. This could be interpreted as an effective dimension $D = 1.52$ which interpolates the 1D initial transformation dynamics and the true 2D process at the late stage when the transformed metallic stripes significantly overlap with each other.

\section{summary and discussion}
\label{sec:summary}

To summarize, we have presented extensive cell dynamics simulations on the phase transition dynamics of two coupled non-conserved scalar fields $\phi$ and $\psi$, representing the charge and magnetic orders, respectively, in correlated nickelates. The special condition relating the ordering wave vectors $\mathbf k_\phi = 2 \mathbf k_\psi$ allows a unique linear-quadratic coupling between the two order parameters in the associated Landau free energy.  This particular theory can also be applied to describe the coupled charge and spin density waves in the collinear stripe phase of high-temperature cuprates. Our large-scale cell dynamics simulations have uncovered a two-stage phase ordering phenomenon. Immediately after a quench into the low-temperature ordered phase, the two scalar fields evolve independently of each other and both obey the Allen-Cahn $L\sim t^{1/2}$ behavior during the first dynamical regime. This initial stage of independent ordering is followed by a transient re-alignment period in which the charge domains of negative order parameter disappear due to the coupling term that explicitly breaks the $Z_2$ symmetry of charge order. The second dynamical regime describing the long-term coarsening of the magnetic domains again follows the Allen-Cahn power law. We note in passing that the same long-term phase-ordering dynamics has been observed in a similar model C system in which a $Z_2$-symmetric non-conserved order parameter is coupled to a conserved concentration field. 

We have further applied the cell dynamics method to simulate the nucleation and growth phenomena of this system. By properly taking into account the transient nucleation effects and finite size of critical nuclei, we show that the kinetics of transformation to the ordered states is well described by the two-dimensional Kolmogorov-Johnson-Mehl-Avrami theory with an exponent $n = 3$. More interestingly, the transition to the high-temperature paramagnetic phase exhibits a very different dynamics due to pre-existing magnetic domain-walls in the initial ordered state. Essentially, these domain walls with reduced charge-order serves as seed of the nucleation, and the phase transformation takes place through the decay of such quasi-metallic domain-wall into two insulator-metal boundaries enclosing a paramagnetic metallic region. The transformation dynamics is consistent with an Avrami exponent $n = 1$, indicating the one-dimensional growth nature with pre-existing nuclei. 

The metallic strings observed in the nano-IR experiments on NdNiO$_3$ films~\cite{post18} have been suggested to be associated with the hidden magnetic domain-walls. Our cell-dynamics simulations also confirmed this intriguing observation. The second-order transition that seems to take place concurrently with the first-order insulator-to-metal transition observed in the nano-IR images is correctly attributed to the gradual decrease of charge order in the vicinity of the domain walls~\cite{post18}. Our detailed analysis reveals that these magnetic domains acting as seed of nucleation dramatically modifies the kinetics of phase transformation to the metallic state. We expect this remarkable disparity between transition into the disordered and ordered phases will manifest itself quantitatively in the measured time dependence of the transformed volume fraction during annealing and warming of the sample. 

The presence of quenched disorder, which is not included in our study, is expected to further complicate the above picture. Indeed, statistical analysis of cluster morphology during the phase transformation seems to conform with the universality class of random field Ising model~\cite{post18,perkovic95,liu16}, underscoring the importance of disorder. The effects of disorder can also be seen in the appearance of large metallic ribbon-like structures that survive down to relatively low temperatures during annealing. Quenched disorder, such as lattice defects, can act as pinning centers for the magnetic domains. More importantly, these impurities can themselves be the seed of thermal induced nucleation. Consequently, the transformation dynamics to the metallic phase might exhibit mixture of 1D domain-wall expansion and 2D nucleation and growth. We expect detailed cell dynamics simulation taking into account the quenched disorder can shed light on the insulator-to-metal transition kinetics and the connection with the random-field Ising universality class. 

\bigskip

\begin{acknowledgments}
{\em Acknowledgements.} G.-W. Chern is grateful to S.~A.~Egorov for insightful discussion. This work is supported by the US Department of Energy Basic Energy Sciences under Contract No. DE-SC0020330. The author also acknowledge the support of Advanced Research Computing Services at the University of Virginia.
\end{acknowledgments}

\newpage

\end{document}